\documentclass[aps,pra,twocolumn,superscriptaddress]{revtex4-1}
\usepackage{bm}
\usepackage{graphicx}
\usepackage{amssymb,amsmath,amsbsy,amsgen,amsfonts}
\usepackage{dcolumn}
\usepackage{amsthm}
\usepackage{mathrsfs}
\usepackage{latexsym}
\usepackage{array}
\usepackage{amstext}
\usepackage{epsfig}
\usepackage{epstopdf} 

\usepackage{color}
\usepackage{units}
\usepackage{calrsfs}
\usepackage{verbatim}
\DeclareMathAlphabet\mathbfcal{OMS}{cmsy}{b}{n}

\newcommand{\be}{\begin{equation}}
\newcommand{\ee}{\end{equation}}
\newcommand{\ba}{\begin{array}}
\newcommand{\ea}{\end{array}}
\newcommand{\bqa}{\begin{eqnarray}}
\newcommand{\eqa}{\end{eqnarray}}

\begin{document}

\title{Non-Markovian Transient Casimir-Polder force and population dynamics on excited and ground state atoms: weak and strong coupling regimes in generally non-reciprocal environments}

\author{George W. Hanson}
\email{george@uwm.edu}
\address{Department of Electrical Engineering, University of Wisconsin-Milwaukee, 3200 N. Cramer St., Milwaukee, Wisconsin 53211, USA}

\author{S. Ali Hassani Gangaraj}
\email{ali.gangaraj@gmail.com}
\address{School of Electrical and Computer Engineering, Cornell University, Ithaca, NY 14853, USA}

\author{ M\'ario G. Silveirinha}
\email{mario.silveirinha@co.it.pt}
\address{Instituto Superior T\'{e}cnico, University of Lisbon
	and Instituto de Telecomunica\c{c}\~{o}es, Torre Norte, Av. Rovisco
	Pais 1, Lisbon 1049-001, Portugal}

\author{Mauro Antezza}
\email{mauro.antezza@umontpellier.fr}
\address{Laboratoire Charles Coulomb (L2C), UMR 5221 CNRS-Universit\'{e} de Montpellier, F-34095 Montpellier, France}
\address{Institut Universitaire de France, 1 rue Descartes, F-75231 Paris Cedex 05, France}

\author{Francesco Monticone}
\email{francesco.monticone@cornell.edu}
\address{School of Electrical and Computer Engineering, Cornell University, Ithaca, NY 14853, USA}

\date{\today}

\begin{abstract}

The transient Casimir-Polder force on a two-level atom introduced into a three-dimensional, inhomogeneous, generally non-reciprocal environment is evaluated using non-Markovian
Weisskopf-Wigner theory in the strong and weak coupling regimes. Ground-state and excited atoms are considered as two
separate initial-value problems, and both the short-time and long-time atomic
population and force are evaluated. The results are compared with various Markov
approximation of the Weisskopf-Wigner theory, and with previous Markov results
from the Heisenberg picture.

\end{abstract}

\maketitle


\section{Introduction}

The population and vacuum forces on atoms (real or artificial ones) is of fundamental interest, and important for practical applications
in atomic control \cite{Nobel1}-\cite{Nobel3} and quantum information \cite{QI}. Particularly for neutral atoms, vacuum forces \cite{CasimirPolder}-\cite{Mauro} and population-related spontaneous-emission effects play an important role.

In an inhomogeneous environment, spontaneous emission can exert a force on atomic systems. In previous work \cite{PRA2018}-\cite{QT}, the quantum force and torque on an excited two-level atom in a non-reciprocal environment (a biased plasma interface) was modeled using the Heisenberg picture. It was found that even in a translationally-invariant environment, a lateral force can exist due to the non-reciprocal nature of the surface plasmon polaritons (SPPs). The analysis in \cite{PRA2018}-\cite{QT} was based on a Markovian solution of the
Heisenberg equations of motion (HEM). The Markov approximation (MA), in conjunction with the Sokhotski--Plemelj (SP) identity, allowed the identification of both resonant and non-resonant force contributions \cite{OptF1}-\cite{OptF2}. In the
limit $t\rightarrow\infty$, the non-resonant force was shown to be equal to the usual Casimir-Polder (CP) force, which is vertically-directed with
respect to the interface. The case of short time dynamics is more delicate, and the previous paper \cite{PRA2018} has shown that the MA, together with the use of the SP identity, leads to a non-zero force at the start of the time origin, $t=0$.

In this work, the correct short-time (transient) behavior of the Casimir-Polder force is determined by removing the Markov approximation, and, in particular, avoiding the use of the SP identity. The atom, introduced into an environment at $t=0$, dynamically self-dresses even for a ground-state atom, because its initial state, although an eigenstate of the unperturbed Hamiltonian, is not an eigenstate of the interacting Hamiltonian \cite{DCE1}. For weak coupling, it is found that the fundamental Markov approximation can lead to reasonable results with the correct force behavior near the time origin, although other commonly-used approximations that enable use of the SP identity lead to incorrect short-time-behavior. 

While there has been a large number of studies on the static CP force (see, e.g., \cite{B1}-\cite{B2} and references therein), studies of the transient CP force have been limited in scope (e.g., in \cite{B2} a Jaynes-Cummings, single mode field is assumed in the strong coupling case). And, there have been few studies of the non-Markovian CP force \cite{NM}. In this work, we consider the initial-value problem of introducing either an excited or a ground state atom into an environment at $t=0$, which is a special case of the Dynamical Casimir Effect, which also encompasses photon generation from fast changes in geometry \cite{DCE}. This problem was considered in \cite{DCE2}-\cite{DCE3} using an expansion of the Heisenberg equation of motion. Here, we use the Weisskopf-Wigner method, applicable to both weak and strong coupling regimes, and which rigorously includes non-Markovian effects. We also extend the formulation to non-reciprocal materials (nonreciprocal continuum reservoir), although non-reciprocity is not needed for the studied effects.

We work within the Schr\"{o}dinger picture, which necessitates
elucidating the joint atomic-field states, and results in treating the
excited-atom and ground-state atom as independent initial-value problems,
since the respective states evolve independently. Regarding
the Casimir-Polder force on a ground-state atom, we show that it arises from
non-energy-conserving states. Some parameters are identified to assess the
strength of the non-Markovian aspect of the response. The formulation is made
for generally nonreciprocal environments, in part to make contact with the
work in \cite{PRA2018}-\cite{QT}, and for applications related to photonic topological insulators, although the main ideas are general and do not necessitate having a nonreciprocal environment.

We now provide a brief comparison of the HEM and Schr\"{o}dinger
Picture methods, in order to clarify the various approximations used. Both
start from the same Hamiltonian. In the HEM, the time-evolution of the atomic
and field operators is derived as a coupled set of equations from the
Heisenberg equation of motion. Solution of the resulting coupled set of equations is
extremely difficult, although the field operator equation can be solved by making a
one-excitation approximation \cite{SH1}. However, as this eliminates
higher-order correlations, more typically a Markov approximation is made,
wherein the dipole operator is assumed to be memoryless. Usually, then, the
upper time-limit of the spectral integral is approximated as $t\rightarrow
\infty$, and the SP identity leads to resonant and non-resonant terms,
the latter being a principal-value integral associated with an energy shift of
the atomic transition. In \cite{PRA2018}, we then wrote both contributions in
terms of the system Green function, which allows complicated environments (e.g., lossy, inhomogeneous, nonreciprocal)
acting as reservoirs to be modeled exactly, in a macroscopic sense.  Alternatively, in this work we use the Weisskopf-Wigner method \cite{WW}-\cite{LMS}, which can also incorporate the Green function. In this case, the MA, although also
widely-used, is not necessary, and the exact solution can be obtained numerically
by solving a Volterra integral equation of the second kind. This leads to the
non-Markovian (non-exponential) evolution of the population, which can be used
in evaluating the exact dipole force. Various MA-type approximations can also
be used in approximating the force, and are discussed in several appendices. 

One complication of the Weisskopf-Wigner method is that atom-field product states need to be defined. Considering a two-state atom defining a two-dimensional Hilbert space $H_{a}=\{e,g\}$, and multimode field Fock states $\{0,1,2,...\}$ defining an infinite-dimensional Hilbert space $H_{f}$, where $0,1,2,...$ represent the number of quanta in a generic field mode, the product states $H_{a} \otimes H_{f}$ separate into two groups, $A=\{\vert
e,0 \rangle, \vert g,1 \rangle, \vert e,2 \rangle, \vert g,3 \rangle, \vert e,4 \rangle,...\}  $ and $B=\{  \vert g,0\rangle, \vert e,1\rangle, \vert g,2\rangle, \vert e,3\rangle, \vert g,4\rangle,...\}$ that evolve independently. An initially-excited atom evolves within Group A, and, hence, cannot decay into the ground state of the non-interacting system $\vert g,0\rangle $.
That is, in the final state the atom can be in the ground state, but the field will have one or more excitations (even in the lossy case). However,
the evolution of the non-interacting system ground state can also be determined, where, even starting from the state $\vert g,0\rangle$, there is
population evolution and force since the direct-product state is not an eigenstate of the full Hamiltonian (except at $t=0$, assuming that the interaction is switched on at that time). Thus, the initially-excited atom case and the ground-state atom case need to be treated
as two independent initial-value problems, which is not necessary with the HEM method.

The article is organized as follows. In Section \ref{NSR}, we describe the generally inhomogeneous, nonreciprocal environment (i.e., a structured reservoir) into which an excited or ground-state atom is introduced. In Section \ref{IEA}, we consider introducing an excited atom into the structured reservoir at $t=0$, and we solve for the non-Markovian atomic population in terms of a Volterra integral equation (VIE) of the second kind. We show that the structural form of the VIE is the same as in the reciprocal case, obtained previously, with non-reciprocity simply entering via the Green function. A new expression for the non-Markovian force dynamics is then obtained, and applied to both weak and strong coupling regimes. In particular, transient force dynamics are studied, where it is shown that the force is initially repulsive, and then oscillates in sign before settling down to become its static attractive value. For strong coupling to a multimode reservoir, we show Rabi oscillations in the force. In Section \ref{GSA}, we repeat the analysis for a ground-state atom, which
leads to the transient Casimir-Polder force, also a new result, exhibiting Rabi oscillations in the strong coupling regime. Finally, we obtain the long-time dynamics using Laplace transforms, and obtain expressions involving a
parameter that indicates the degree of non-Markovian behavior. After some some concluding remarks, appendices provides details of the numerical method used to solve the Volterra integral equation, and several different
Markov-type approximations of the population and force.


\section{Nonreciprocal Structured Reservoir Environment\label{NSR}}

In a translationally-invariant and reciprocal environment, spontaneous emission occurs randomly in all directions, so that the net force on a linearly polarized, initially-excited atom is zero. For an atom near an interface, the Casimir-Polder force is present, associated with vacuum fluctuations and the change of the photonic density of states brought about by the presence of the interface. In addition to the force perpendicular to the interface, as shown in \cite{PRA2018}-\cite{PRB2018}, at
an interface between a nonreciprocal medium and a simple medium, unidirectional surface plasmon polaritons mediate non-null lateral spontaneous emission forces. 

In the following, we consider introducing an excited-state or
ground-state atom at $t=0$ into a lossy, inhomogeneous, and non-reciprocal environment, which serves as a structured reservoir for the atom, and examine the time-dynamics of the resulting atomic population, spontaneous emission (SE) rate, and force. The problem is cast as an initial-value problem using Weisskopf-Wigner theory \cite{WW}-\cite{LMS}, adapted for the non-reciprocal medium.

\begin{figure}[!htbp]
	\begin{center}
		\noindent \includegraphics[width=2.1in]{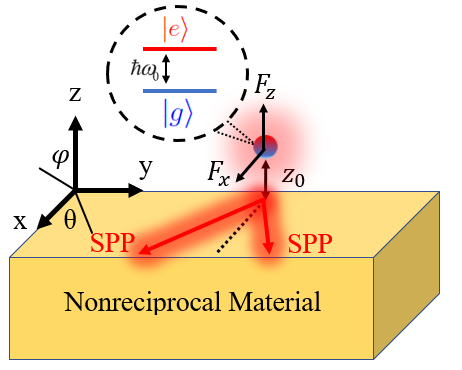}
	\end{center}
	\caption{A two-level system near the surface of a nonreciprocal material, experiencing the Casimir-Polder force, and, for an excited atom, a spontaneous emission optical force. The main decay channel is the SPPs on the interface.}
	\label{Fig_geom}
\end{figure}

Figure \ref{Fig_geom} depicts the situation, where an two-level atom resides in the
vicinity of a material interface.%

We suppose the region $z>0$ is filled by vacuum, and that the region $z<0$ is
filled with a gyrotropic material with permittivity ${\boldsymbol{\varepsilon
}}=\varepsilon_{0}(\varepsilon_{t}{\boldsymbol{\mathrm{I}}}_{t}+\varepsilon
_{a}\mathbf{{\hat{y}}}\mathbf{{\hat{y}}}+i\varepsilon_{g}\mathbf{{\hat{y}}%
}\times\boldsymbol{\mathrm{I}})$, where ${\boldsymbol{\mathrm{I}}}%
_{t}=\boldsymbol{\mathrm{I}}-\mathbf{{\hat{y}}{\hat{y}}}$, with $\varepsilon
_{g}$ being the magnitude of the gyration pseudovector. For the gyrotropic
medium we consider a magnetized plasma (e.g., InSb \cite{Palik}). For a static
bias magnetic field along the $+y$-axis the permittivity components are
\cite{Bittencourt}
\begin{align}
&  {\varepsilon_{t}}=1-\frac{{\omega_{p}^{2}\left(  {1+i\Gamma/\omega}\right)
}}{{{{\left(  {\omega+i\Gamma}\right)  }^{2}}-\omega_{c}^{2}}}%
\nonumber\label{bp}\\
&  {\varepsilon_{a}}=1-\frac{{\omega_{p}^{2}}}{{\omega\left(  {\omega+i\Gamma
}\right)  }},\,\,\,\,{\varepsilon_{g}}=\frac{1}{\omega}\frac{{\omega_{c}%
\omega_{p}^{2}}}{{\omega_{c}^{2}-{{\left(  {\omega+i\Gamma}\right)  }^{2}}}}.
\end{align}
Here, $\omega_{p}$ is the plasma frequency, $\Gamma$ is the collision rate
associated with damping, $\omega_{c}=-qB_{0}/m>0$ is the cyclotron frequency,
$q=-e$ is the electron charge, $m$ is the electron effective mass, and $B_{0}$
is the static bias. In the special case that $B_{0}=0$, the system is
reciprocal. A limitingly-lowloss plasma is assumed for simplicity, since loss does not qualitatively affect the time-dynamics of interest. The analytical form of the
Green function for this environment is provided in \cite{PRA2018}.

In the following, we
assume that the dipole is linearly-polarized, $\mathbf{\gamma}=\widehat
{\mathbf{z}}\gamma_{z}$, with $\gamma_{z}$ real-valued, located a distance $z_{0}$ from the
interface, and we take $\omega_{p}= (2 \pi) 200 \times 10^{12}$ Hz and $\omega
_{0}=0.65\omega_{p}$. 

\section{Initially-Excited Atom Introduced into a Non-Reciprocal Structured
Reservoir\label{IEA}}

In this section, we consider introducing an excited-state atom at $t=0$ into
the structured reservoir described above. The ground-state atom is considered
in Section \ref{GSA}.

\subsection{Initially-Excited Atom: Schr\"{o}dinger Picture Wavefunction
Amplitude Evolution in a Non-Reciprocal Environment\label{WAS}}

In the Schr\"{o}dinger picture, the system Hamiltonian is
\cite{QED}%
\begin{align}
\widehat{\mathrm{H}}  =\int d^{3}\mathbf{r}\int_{0}^{\infty}& d\omega
_{\lambda}\hslash\omega_{\lambda}\hat{\boldsymbol{\mathrm{f}}}^{\dagger
}(\mathbf{r},\omega_{\lambda})\hat{\boldsymbol{\mathrm{f}}}(\mathbf{r}%
,\omega_{\lambda})\label{Ham}\\
&  +\hbar\omega_{0}\hat{\sigma}_{+}\widehat{{\sigma}}_{-}-\hat{\mathbf{p}%
}\cdot\widehat{\boldsymbol{\mathrm{E}}}(\mathbf{r}_{0}),\nonumber
\end{align}
where the first term is the Hamiltonian for the field modes, the second term
is the Hamiltonian for the atomic operators, and the last term accounts for
the field-atom coupling. In (\ref{Ham}), $\hat{\boldsymbol{\mathrm{f}}}%
,\hat{\boldsymbol{\mathrm{f}}}^{\dag}$ are the canonically conjugate field
variables (continuum bosonic operator--valued vectors of the combined
matter-field system) that satisfy
\begin{align}
\left[  \widehat{f}_{k}\left(  \mathbf{r},\omega\right)  ,\widehat
{f}_{k^{\prime}}^{\dag}\left(  \mathbf{r}^{\prime},\omega^{\prime}\right)
\right]   &  =\delta_{kk^{\prime}}\delta\left(  \omega-\omega^{\prime}\right)
\delta\left(  \mathbf{r}-\mathbf{r}^{\prime}\right)  ,\\
\left[  \widehat{f}_{k}\left(  \mathbf{r},\omega\right)  ,\widehat
{f}_{k^{\prime}}\left(  \mathbf{r}^{\prime},\omega^{\prime}\right)  \right]
&  =\left[  \widehat{f}_{k}^{\dag}\left(  \mathbf{r},\omega\right)
,\widehat{f}_{k^{\prime}}^{\dag}\left(  \mathbf{r}^{\prime},\omega^{\prime
}\right)  \right]  =0,
\end{align}
$\hat{\sigma}_{\pm}$ are the canonically conjugate two-level atomic operators
($\widehat{\sigma}_{+}=\left\vert e\right\rangle \left\langle g\right\vert
,\ \widehat{\sigma}_{-}=\left\vert g\right\rangle \left\langle e\right\vert
=\widehat{\sigma}_{+}^{\dagger}$, with $\left\vert e\right\rangle $ and
$\left\vert g\right\rangle $ being the excited and ground atomic states,
respectively), and $\hat{\mathbf{p}}=\left(  \hat{\sigma}_{+}+\hat{\sigma}%
_{-}\right)  \mathbf{\gamma}$ is the dipole operator, where $\mathbf{\gamma}$
is the dipole operator matrix-element.

For the atom-field system, we define product states such as $\left\vert
e,0\right\rangle \equiv\left\vert e\right\rangle \otimes\left\vert \left\{
0\right\}  \right\rangle $ and\ $\left\vert g,1_{i}\left(  \mathbf{r}%
,\omega_{\lambda}\right)  \right\rangle \equiv\left\vert g\right\rangle
\otimes\left\vert \left\{  1_{i}\left(  \mathbf{r},\omega_{\lambda}\right)
\right\}  \right\rangle $. The state $\left\vert 1_{i}\left(  \mathbf{r}%
,\omega_{\lambda}\right)  \right\rangle =\left\vert \left\{  1_{i}\left(
\mathbf{r},\omega_{\lambda}\right)  \right\}  \right\rangle $ indicates that
the $\lambda^{\text{th}}$ field mode of the nonuniform continuum is populated with a single quanta, and that
it is vector-valued with field component in the $i^{\text{th}}$ direction. It
can be noted that if one uses, rather than the full interaction Hamiltonian
$\hat{\mathbf{p}}\cdot\widehat{\boldsymbol{\mathrm{E}}}(\mathbf{r}_{0})$, the
rotating wave approximation (RWA) interaction Hamiltonian which contains
$\left(  \widehat{\sigma}_{+}\hat{\boldsymbol{\mathrm{f}}}+\text{H.c.}\right)
$, then the initial state $\left\vert e,0\right\rangle $ produces only
$\left\vert g,1\right\rangle $. However, the full interaction Hamiltonian
$\hat{\mathbf{p}}\cdot\widehat{\boldsymbol{\mathrm{E}}}(\mathbf{r}_{0}%
)\sim\left(  \hat{\sigma}_{+}+\hat{\sigma}_{-}\right)  \left(  \hat
{\boldsymbol{\mathrm{f}}}+\hat{\boldsymbol{\mathrm{f}}}^{\dagger}\right)  $
acting on the initial state $\left\vert e,0\right\rangle $ leads to an
infinite-dimensional Hilbert space of the set of states $A=\left\{  \left\vert
e,0\right\rangle ,\left\vert g,1\right\rangle ,\left\vert e,2\right\rangle
,\left\vert g,3\right\rangle ,\left\vert e,4\right\rangle ,...\right\}  $,
where the $n>1$ photons could be in the same or different field modes. For the excited atom,
we truncate the space to consist of \{$\left\vert e,0\right\rangle ,\left\vert
g,1\right\rangle $\}, which is equivalent to a rotating wave approximation
even when using the full interaction Hamiltonian. Later, we consider
non-energy-conserving states, which are necessary for the analysis of the
ground-state atom.

We assume a general inhomogeneous, lossy, and non-reciprocal environment
characterized by the permittivity tensor $\mathbf{\varepsilon}\left(
\mathbf{r},\omega\right)  $. We follow the phenomenological
macroscopic Langevin noise approach \cite{Welsch0}-\cite{Hanson} (see also
\cite{AD}, where a comparison with a generalized Huttner-Barnett approach is
discussed, and also \cite{Phil}, where the phenomenological assumptions are
derived from a canonical formulation). The quantized Schr\"{o}dinger picture
electric field operator is%
\begin{align}
\widehat{\mathbf{E}}\left(  \mathbf{r}\right)  =\int_{0}^{\infty}%
d\omega_{\lambda}\ \  &  i\sqrt{\frac{\hslash}{\pi\varepsilon_{0}}}%
\frac{\omega_{\lambda}^{2}}{c^{2}}\int d^{3}\mathbf{r}^{\prime}%
\boldsymbol{\mathrm{G}}(\mathbf{r},\mathbf{r}^{\prime},\omega_{\lambda
})\label{E}\\
&  \cdot\boldsymbol{\mathrm{T}}(\mathbf{r}^{\prime},\omega_{\lambda}%
)\cdot\widehat{\mathbf{f}}\left(  \mathbf{r}^{\prime},\omega_{\lambda}\right)
+\text{H.c.}\nonumber
\end{align}
where $\boldsymbol{\mathrm{T}}(\mathbf{r},\omega_{\lambda})\cdot
\boldsymbol{\mathrm{T}}^{\dag}(\mathbf{r},\omega_{\lambda})=\frac{1}%
{2i}\left(  \mathbf{\varepsilon}\left(  \mathbf{r},\omega\right)
-\mathbf{\varepsilon}^{\dag}\left(  \mathbf{r},\omega\right)  \right)  $; for
reciprocal media, $\mathbf{T}=\sqrt{\operatorname{Im}\left\{  \varepsilon
\left(  \mathbf{r},\omega\right)  \right\}  }\mathbf{I}$, and where
$\boldsymbol{\mathrm{G}}(\mathbf{r},\mathbf{r}^{\prime},\omega_{\lambda})$ is
the classical Green function for the nonreciprocal environment, discussed in Appendix \ref{A_GF}. We
assume that an atom is introduced to the environment at $t=0$. Furthermore, we assume zero temperature, and that the atomic
transition frequency $\omega_{0}$ is not too close to a material
resonance. Otherwise, there could be additional
transients \cite{Wubs} that are ignored here. 

The equation of motion (Schr\"{o}dinger equation) is $\left(  d/dt\right)
\left\vert \psi\right\rangle =-\left(  i/\hslash\right)  \widehat{H}\left\vert
\psi\right\rangle $. Using the energy-conserving states (ECS) \{$\left\vert
e,0\right\rangle ,\left\vert g,1_{i}\left(  \mathbf{r},\omega_{\lambda
}\right)  \right\rangle $\}, the expansion of the wavefunction is
\begin{align}
\left\vert \psi\left(  t\right)  \right\rangle _{\text{ECS}} &  =b_{eo}\left(
t\right)  \left\vert e,0\right\rangle \\
&  +\int d^{3}\mathbf{r}\int_{0}^{\infty}d\omega_{\lambda}b_{g1i}\left(
\mathbf{r},\omega_{\lambda},t\right)  \left\vert g,1_{i}\left(  \mathbf{r}%
,\omega_{\lambda}\right)  \right\rangle ,\nonumber
\end{align}
where $b_{eo}\left(  t\right)  $ is the atomic excited state population
amplitude. Here and in the following we sum over repeated vector-component
indices. Conservation of probability requires%
\begin{equation}
\left\vert b_{eo}\left(  t\right)  \right\vert ^{2}+\int_{0}^{\infty}%
d\omega_{\lambda}\int d\mathbf{r}\left\vert b_{gi}\left(  \mathbf{r}%
,\omega_{\lambda},t\right)  \right\vert ^{2}=1.\label{CP}%
\end{equation}

It is convenient to write $b_{eo}\left(  t\right)  =c_{eo}\left(  t\right)
e^{-i\omega_{0}t}$ and $b_{g1i}\left(  \mathbf{r},\omega_{\lambda},t\right)
=c_{g1i}\left(  \mathbf{r},\omega_{\lambda},t\right)  e^{-i\omega_{\lambda}t}%
$. Plugging the wavefunction into the Schr\"{o}dinger equation and using
orthogonality, for $\mathbf{\gamma}=\widehat{\mathbf{x}}_{j}\gamma_{j}$, it is
straightforward to obtain the coupled set of equations ($j$ is fixed)
\begin{align}
&  \frac{d}{dt}c_{eo}\left(  t\right)  =-\gamma_{j}\sqrt{\frac{1}{\hslash
\pi\varepsilon_{0}}}\int_{0}^{\infty}d\omega_{\lambda}\ \frac{\omega_{\lambda
}^{2}}{c^{2}}\label{ce}\\
&  \times\int d^{3}\mathbf{r}^{\prime}K_{ji}\left(  \mathbf{r}_{0}%
,\mathbf{r}^{\prime},\omega_{\lambda}\right)  c_{g1i}\left(  \mathbf{r}%
^{\prime},\omega_{\lambda},t\right)  e^{-i\left(  \omega_{\lambda}-\omega
_{0}\right)  t},\nonumber\\
&  \frac{d}{dt}c_{g1i}\left(  \mathbf{r},\omega_{\lambda},t\right)
=\sqrt{\frac{1}{\hslash\pi\varepsilon_{0}}}\gamma_{j}\frac{\omega_{\lambda
}^{2}}{c^{2}}K_{ji}^{\ast}\left(  \mathbf{r}_{0},\mathbf{r},\omega_{\lambda
}\right)  \label{cg}\\
&
\ \ \ \ \ \ \ \ \ \ \ \ \ \ \ \ \ \ \ \ \ \ \ \ \ \ \ \ \ \ \ \ \ \ \ \ \ \ \ \times
c_{eo}\left(  t\right)  e^{i\left(  \omega_{\lambda}-\omega_{0}\right)
t},\nonumber
\end{align}
where $\mathbf{K}\left(  \mathbf{r},\mathbf{r}^{\prime},\omega_{\lambda}\right)
=\boldsymbol{\mathrm{G}}(\mathbf{r},\mathbf{r}^{\prime},\omega_{\lambda}%
)\cdot\boldsymbol{\mathrm{T}}(\mathbf{r}^{\prime},\omega_{\lambda}).$ It can be noted that (\ref{ce})-(\ref{cg}) are the same as 
\cite[(6.26)-(6.27)]{QED}\ and \cite[(23)-(24)]{SD}, except here generalized to
nonreciprocal media.

Integrating (\ref{cg}), assuming that the excitation initially resides in the
atom, $c_{g1i}\left(  \mathbf{r},\omega_{\lambda},t=0\right)  =0$, and inserting
the result into (\ref{ce}) and using (\ref{magic}) leads to the non-Markovian population equation
in the form of a Volterra integral equation of the second kind,
\begin{equation}
\frac{d}{dt}c_{eo}\left(  t\right)  =\int_{0}^{t}H\left(  t,t^{\prime}\right)
c_{eo}\left(  t^{\prime}\right)  dt^{\prime},\label{VE1}%
\end{equation}
with the kernel%
\begin{align}
H\left(  t,t^{\prime}\right)  =-\frac{1}{\hslash\pi\varepsilon_{0}}\int
_{0}^{\infty}d\omega_{\lambda}\ \frac{\omega_{\lambda}^{2}}{c^{2}} &
\frac{\mathbf{\gamma}\cdot\mathbf{G}_{\mathbf{I}}(\mathbf{r}_{0}%
,\mathbf{r}_{0},\omega_{\lambda})\cdot\mathbf{\gamma}}{2i}\nonumber\\
&  \times e^{-i\left(  \omega_{\lambda}-\omega_{0}\right)  \left(
t-t^{\prime}\right)  } \label{ker1},
\end{align}
where $G_{\mathbf{I},i,j}\left(
\mathbf{r},\mathbf{r}_{0},\omega_{\lambda}\right)=G_{ij}(\mathbf{r},\mathbf{r}_{0},\omega_{\lambda})-G_{ij}^{\ast
}(\mathbf{r}_{0},\mathbf{r},\omega_{\lambda}) $ \cite{Buh}. We will assume the initial-value condition $c_{eo}\left(  0\right)  =1$. It is
useful to note that for a linearly-polarized, real-valued dipole moment (assumed here), $\mathbf{\gamma
}\cdot\mathbf{G}_{\mathbf{I}}(\mathbf{r}_{0},\mathbf{r}_{0},\omega_{\lambda
})\cdot\mathbf{\gamma}$ picks out a diagonal element of the Green function,
and $G_{\mathbf{I},ii}(\mathbf{r}_{0},\mathbf{r}_{0},\omega
)=2i\operatorname{Im}G_{ii}(\mathbf{r}_{0},\mathbf{r}_{0},\omega)$, even for a
nonreciprocal medium, and so in that case the form of the Volterra equation
(\ref{VE1}) is the same in the reciprocal and non-reciprocal cases.\textbf{\ }%
The procedure for numerically solving the Volterra integral equation is shown in Appendix \ref{NS}. Appendix \ref{MAEAP} details various levels of Markov approximations that enable closed-form solutions. Specifically, if the population is assumed to be memoryless (Markov approximation (MA)), $c_{eo}\left(  t^{\prime}\right)  \simeq c_{eo}\left(
t\right)  $, the upper limit of the time integral is extended to infinity, and the Sokhotski--Plemelj (SP) identity (\ref{SPI}) is used, we call this the full Markov (FM) approximation. If, however, the MA is made, but the upper
limit of the integration is not extended to infinity, we call this the partial Markov (PM) approximation.

\begin{figure}[!htbp]
	\begin{center}
		\noindent \includegraphics[width=3.5in]{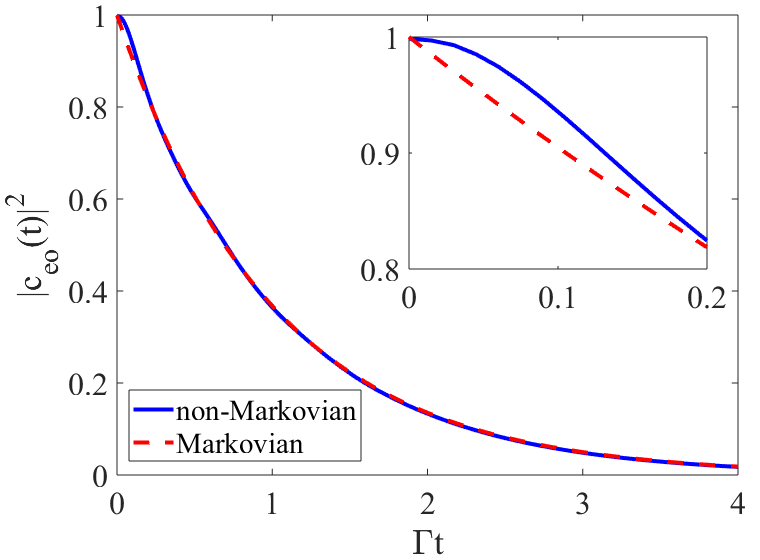}
	\end{center}
	\caption{Non-Markovian population dynamics in the weak-coupling regime, obtained from the numerical solution of (\ref{VE1}), and compared with the usual Markov population decay using
(\ref{beoFM}). The insert shows the behavior near $t=0$. The atom is located $z_0=0.7 c/\omega_p$ above the interface, such that $g=0.044$.}
	\label{fig_pop}
\end{figure}

The coupling parameter
\begin{equation}
g=\frac{\mid \mathbf{\gamma} \mid} {\hslash\omega_{\text{SPP}}} \sqrt{\frac{\hslash\omega
_{\text{SPP}}}{32\pi\varepsilon_{0}z_{0}^{3}}},%
\end{equation}
where we assume $\omega_{\text{SPP}} \approx \omega_0$, delineates weak ($g\ll1$) and strong ($g\geq0.5$) coupling. Figure \ref{fig_pop} shows the non-Markovian population dynamics obtained from the numerical
solution of (\ref{VE1}) for a dipole positioned $z_0=0.7 c/\omega_p$ above the interface, such that  $g=0.044$ indicates weak coupling. Comparison is made to the FM approximation (\ref{beoFM}) (the PM approximation, Eq. (\ref{sPM1}), yields similar results). The non-Markovian result shows the correct zero slope at $t=0$ \cite{FGR1978}-\cite{BF2}, as shown in the insert of  Fig. \ref{fig_pop}. Other then the initial slope, it
can be seen that excellent agreement between the Markov approximation and the non-Markov solution is obtained, as expected for weak coupling. Although not shown, the
non-Markovian solution is also expected to show slower than exponential decay for long times \cite{Mil}.

\begin{figure}[!htbp]
	\begin{center}
		\noindent \includegraphics[width=3.5in]{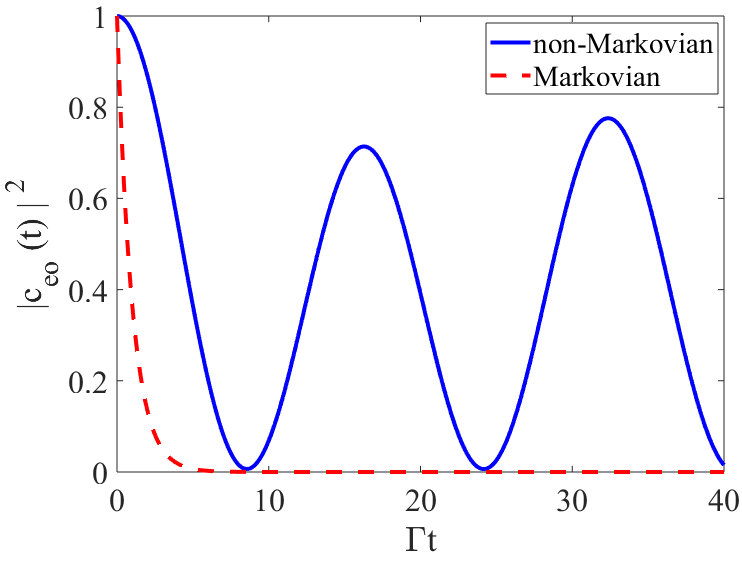}
	\end{center}
	\caption{Non-Markovian population dynamics in the strong-coupling regime, obtained from the numerical solution of (\ref{VE1}), and compared with the usual Markov population decay using
(\ref{beoFM}). The atom is located $z_0=0.1 c/\omega_p$ above the interface, such that $g=0.808$.}
	\label{fig_pop_sc}
\end{figure}

Figure \ref{fig_pop_sc} shows the same result as Fig. \ref{fig_pop}, except for atom height $z_0=0.1 c/\omega_p$ above the interface. In this case, $g=0.808$, indicating strong coupling. The exact solution is strongly non-Markovian, as expected, and exhibits Rabi oscillations. 

\subsection{Initially-Excited Atom: Transient Non-Markovian Casimir-Polder Force in a Non-Reciprocal Environment}

From canonical quantization, the quantum operator for the dipole force on an
atom located at $\mathbf{r}_{0}$ is \cite{CATsm}
\begin{equation}
\hat{\mathcal{F}}_{j}=\hat{\mathbf{p}}\cdot\left.  \frac{\partial}{\partial
j}\hat{\mathbf{E}}\left(  \mathbf{r}\right)  \right\vert _{\mathbf{r}%
=\mathbf{r}_{0}},~~~j=x,y,z.\label{F}%
\end{equation}
The expectation value of force operator in the $\alpha$th direction due to a
dipole oriented along the $j$th coordinate is
\begin{align}
\mathcal{F}_{\alpha}^{j}\left(  t\right)   &  =\left\langle \hat
{\mathcal{F}_{\alpha}}^{j}\right\rangle \\
&  =\left\langle \psi\left(  t\right)  \right\vert \left(  \left(  \hat
{\sigma}_{+}+\hat{\sigma}_{-}\right)  \gamma_{j}\widehat{\mathbf{x}}_{j}%
\cdot\partial_{\alpha}\left.  \widehat{\mathbf{E}}\left(  \mathbf{r}\right)
\right\vert _{\mathbf{r}=\mathbf{r}_{0}}\right)  \left\vert \psi\left(
t\right).  \right\rangle \notag
\end{align}
Using (\ref{cg}), (\ref{magic}), and summing over repeated indices, the
general non-Markovian force is%
\begin{widetext}
\begin{equation}
\mathcal{F}_{\alpha}\left(  t\right)  =2\operatorname{Re}\left\{  \frac{i}%
{\pi\varepsilon_{0}}c_{eo}^{\ast}\left(  t\right)  \int_{0}^{\infty}%
d\omega_{\lambda}\frac{\omega_{\lambda}^{2}}{c^{2}}\left.  \frac{\partial
}{\partial\alpha}\frac{\mathbf{\gamma}\cdot\mathbf{G}_{\mathbf{I}}\left(
\mathbf{r},\mathbf{r}_{0},\omega_{\lambda}\right)  \cdot\mathbf{\gamma}}%
{2i}\right\vert _{\mathbf{r}=\mathbf{r}_{0}}\int_{0}^{t}c_{eo}\left(
t^{\prime}\right)  e^{-i\left(  \omega_{\lambda}-\omega_{0}\right)  \left(
t-t^{\prime}\right)  }dt^{\prime}\right\} \label{F2}%
\end{equation}
\end{widetext}
for $\alpha=x,y,z$. This is the first main analytical result of this paper. Various Markov approximations of the force are provided in Appendix \ref{MAEA}. In particular, one can substitute the FM or PM
approximations for the population into the force expression, then evaluate the resulting time-integral exactly, leading to what we refer to as the FM or PM approximation, respectively, of the force. Alternatively, one could impose the Markov approximation $c_{eo}\left(t^{\prime}\right)  \simeq c_{eo}\left(  t\right)  $ directly in the time integral in the force expression, then either evaluate the resulting time-integral exactly, which we denote as the PM2 approximation, or extend the upper limit of the time integral to infinity and use the Sokhotski--Plemelj identity, which we denote as the FM2 approximation.  

\begin{figure}[!htbp]
	\begin{center}
		\noindent \includegraphics[width=3.5in]{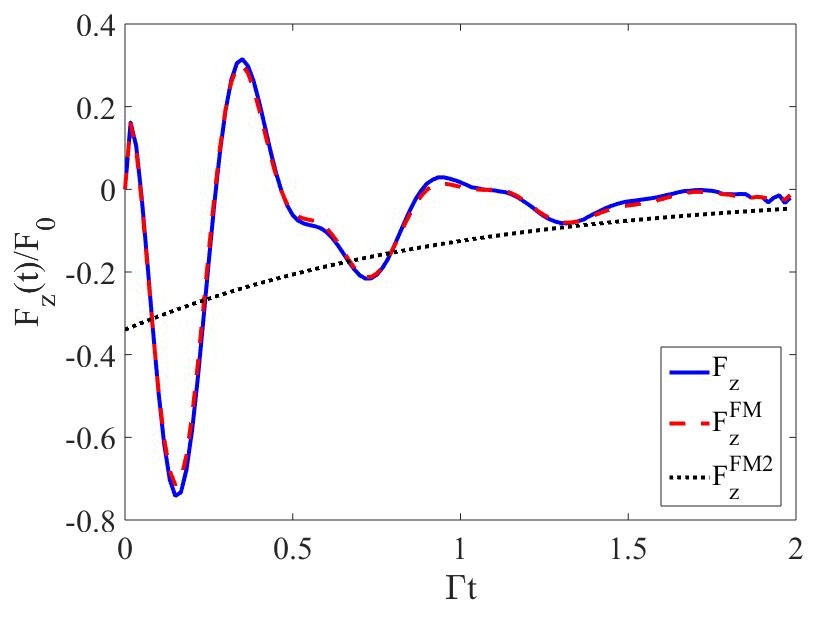}
	\end{center}
	\caption{Normalized non-Markovian vertical force (\ref{F2}) in the weak-coupling regime, compared with the Markov approximation (\ref{FFM}) and the HEM result \cite{PRA2018}. $\mathcal{F}_{0}=3|\mathbf{\gamma}|^{2}/\left(  16\pi z_{0}^{4}\varepsilon_{0}\right)  $ (N). The atom is located $z_0=0.7 c/\omega_p$ above the interface, $g=0.044$.}
	\label{fig_force_vert}
\end{figure}

Figure \ref{fig_force_vert} shows the normalized exact vertical force
$\mathcal{F}_{z}$\ from (\ref{F2}) compared with the FM approximation
(\ref{FFM}), and the result from \cite{PRA2018} which used the Markov approximation of the Heisenberg equations of motion, together with the SP identity, for the weak-coupling case $z_0=0.7 c/\omega_p$. Note that the force is initially repulsive, and then oscillates in sign before settling down to become attractive.

The FM approximation is in good agreement with the exact force (\ref{F2}), indicating that the short-time force dynamics are essentially Markovian in the weak-coupling case.  Importantly, this approximation does not entail use of the SP identity, and has the correct null value at the time origin \cite{FN}. All solutions initially oscillate, and eventually settle-down to the
MA HEM solution, which was obtained in \cite{PRA2018} using the SP identity (which does not provide the correct short-time dynamics). For the nonreciprocal case, a lateral force also exists, but will be omitted here.


\begin{figure}[!htbp]
	\begin{center}
		\noindent \includegraphics[width=3.5in]{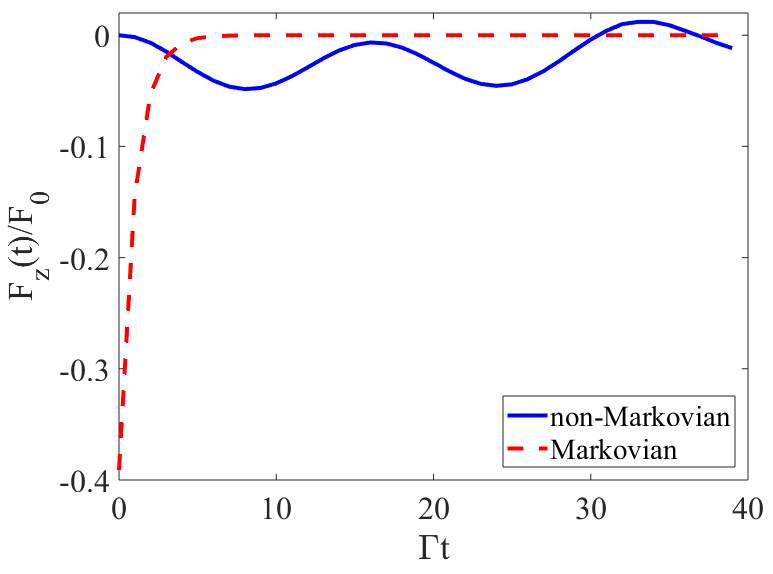}
	\end{center}
	\caption{Normalized non-Markovian vertical force (\ref{F2}) in the strong-coupling regime, compared with the Markov approximation (\ref{FFM}). $\mathcal{F}_{0}=3|\mathbf{\gamma}|^{2}/\left(  16\pi z_{0}^{4}\varepsilon_{0}\right)  $ (N). The atom is located $z_0=0.1 c/\omega_p$ above the interface, $g=0.808$.}
	\label{fig_force_vert_sc}
\end{figure}

Figure \ref{fig_force_vert_sc} shows the normalized exact vertical force
$\mathcal{F}_{z}$\ from (\ref{F2}) compared with the HEM result \cite{PRA2018} for the strong-coupling case, $z_0=0.1 c/\omega_p$. The Rabi oscillations of the population (Fig. \ref{fig_pop_sc}) are evident in the force, indicating strongly non-Markovian behavior.

\section{Casimir-Polder Force on a Ground-State Atom Introduced into a Non-Reciprocal Structured
Reservoir \label{GSA}}

In the Heisenberg picture, atom-field states do not need to be defined, and
the force $\mathcal{F}_{z}$ found via the HEM naturally becomes the
Casimir-Polder force for large times. However, using the Weisskopf-Wigner
method, the states for the excited atom-field are \{$\left\vert
e,0\right\rangle ,\left\vert g,1_{i}\left(  \mathbf{r},\omega_{\lambda
}\right)  \right\rangle $\}, and the joint atom-field ground state is never reached (even using the full set of states $A=\left\{  \left\vert
e,0\right\rangle ,\left\vert g,1\right\rangle ,\left\vert e,2\right\rangle
,\left\vert g,3\right\rangle ,\left\vert e,4\right\rangle ,...\right\}  $). In
this section, we investigate the CP force on a ground-state atom introduced
into a non-reciprocal structured reservoir at $t=0$. We will continue to assume a vertically-polarized atom, although for the ground state a better approximation would be to average over vertical and horizontal polarizations.

\subsection{Ground-State Atom: Non-Markovian Population and Transient Casimir-Polder Force in a
Non-Reciprocal Environment}

When considering the Casimir-Polder force on a ground-state atom, the
assumption is usually that both the atom and field are in the ground state. If we assume an initial state as a direct product of atomic and field ground states, i.e., the non-interacting system ground state $\left\vert g,0\right\rangle $, the full interaction Hamiltonian acts on the initial state to produce the set of states
\{$\left\vert g,0\right\rangle ,\left\vert e,1\right\rangle ,\left\vert
g,2\right\rangle ,\left\vert e,3\right\rangle ,\left\vert g,4\right\rangle
$,...\}, where, again, the numbers represent the number of quanta in the generic field mode. The two sets of states, $A=\left\{  \left\vert e,0\right\rangle
,\left\vert g,1\right\rangle ,\left\vert e,2\right\rangle ,\left\vert
g,3\right\rangle ,\left\vert e,4\right\rangle ,...\right\}  $ used for an
initially-excited atom, and $B=\left\{  \left\vert g,0\right\rangle
,\left\vert e,1\right\rangle ,\left\vert g,2\right\rangle ,\left\vert
e,3\right\rangle ,\left\vert g,4\right\rangle ,...\right\}  $ used for an
initial ground-state atom, are independent (uncoupled). The set $B$ is useful
for the following situation: If we introduce a quasi-ground-state atom $\left\vert g,0\right\rangle $ at $t=0$ into
a structured nonreciprocal reservoir, then the SE and force evolve using set
$B$, in contradistinction to the situation involving an initially-excited atom considered in the previous sections. Here, we truncate the Hilbert space to
consist of the two non-energy-conserving (NEC) virtual states \{$\left\vert
g,0\right\rangle ,\left\vert e,1_{i}\left(  \mathbf{r},\omega_{\lambda
}\right)  \right\rangle $\}, such that the wavefunction is
\begin{align}
\left\vert \psi\left(  t\right)  \right\rangle _{\text{NECS}} &
=b_{go}\left(  t\right)  \left\vert g,0\right\rangle \\
&  +\int d^{3}\mathbf{r}\int_{0}^{\infty}d\omega_{\lambda}b_{e1i}\left(
\mathbf{r},\omega_{\lambda},t\right)  \left\vert e,1_{i}\left(  \mathbf{r}%
,\omega_{\lambda}\right)  \right\rangle .\nonumber
\end{align}
Since the two pairs of states $A$ and $B$ are independent (uncoupled),
$\left\vert \psi\left(  t\right)  \right\rangle _{\text{ECS}}$ and $\left\vert
\psi\left(  t\right)  \right\rangle _{\text{NECS}}$ can be evolved separately.

For the NECS states \{$\left\vert g,0\right\rangle ,\left\vert e,1_{i}\left(
\mathbf{r},\omega_{\lambda}\right)  \right\rangle $\} we find that the
population satisfies the second-kind Volterra integral equation%
\begin{equation}
\frac{d}{dt}b_{go}\left(  t\right)  =\int_{0}^{t}H\left(  t,t^{\prime}\right)
b_{go}\left(  t^{\prime}\right)  dt^{\prime},\label{cg2a}%
\end{equation}
where%
\begin{align}
H\left(  t,t^{\prime}\right)  =-\frac{1}{\hslash\pi\varepsilon_{0}}\int
_{0}^{\infty}d\omega_{\lambda}\ \frac{\omega_{\lambda}^{2}}{c^{2}} &
\frac{\mathbf{\gamma}\cdot\mathbf{G}_{\mathbf{I}}(\mathbf{r}_{0}%
,\mathbf{r}_{0},\omega_{\lambda})\cdot\mathbf{\gamma}}{2i}\nonumber\\
&  \times e^{-i\left(  \omega_{\lambda}+\omega_{0}\right)  \left(
t-t^{\prime}\right)  }. \label{cg2}
\end{align}
assuming $b_{e1i}\left(  \mathbf{r},\omega_{\lambda},t=0\right)  =0$ and the
initial-value condition $b_{go}\left(  0\right)  =1$. Comparing the kernels (\ref{ker1})
and (\ref{cg2}), we see that they are the same except that $\left(
\omega_{\lambda}-\omega_{0}\right)  $ in (\ref{ker1}) is replaced by $\left(
\omega_{\lambda}+\omega_{0}\right)  $ in (\ref{cg2}). Whereas the Markov
approximation of (\ref{VE1})-(\ref{ker1}) leads to both exponential decay and an energy
shift (Section \ref{MAEA}), the Markov approximation of (\ref{cg2a})-(\ref{cg2}) leads to
only an energy shift.

Similar to (\ref{F2}), the non-Markovian force on the ground-state atom is%
\begin{widetext}
\begin{equation}
\mathcal{F}_{\alpha}\left(  t\right)  =2\operatorname{Re}\left\{  \frac{i}%
{\pi\varepsilon_{0}}b_{go}^{\ast}\left(  t\right)  \int_{0}^{\infty}%
d\omega_{\lambda}\frac{\omega_{\lambda}^{2}}{c^{2}}\left.  \frac{\partial
}{\partial\alpha}\frac{\mathbf{\gamma}\cdot\mathbf{G}_{\mathbf{I}}\left(
\mathbf{r},\mathbf{r}_{0},\omega_{\lambda}\right)  \cdot\mathbf{\gamma}}%
{2i}\right\vert _{\mathbf{r}=\mathbf{r}_{0}}\int_{0}^{t}b_{go}\left(
t^{\prime}\right)  e^{-i\left(  \omega_{\lambda}+\omega_{0}\right)  \left(
t-t^{\prime}\right)  }dt^{\prime}\right\} \label{FG}%
\end{equation}
\end{widetext}
for $\alpha=x,y,z$. Together with (\ref{F2}), this is one of the main results of this paper

The non-Markovian population of the ground-state atom is obtained by the numerical solution of
(\ref{cg2a}), using the procedure described in Appendix \ref{NS} (although, due to
the rapidly-oscillating temporal integral in (\ref{cg2a}), a much smaller time
step needs to be used compared to solving (\ref{VE1})). Various Markov approximations of the population and force are provided in Appendices \ref{MAEAP} and \ref{MAEA}, respectively.

\begin{figure}[!htbp]
	\begin{center}
		\noindent \includegraphics[width=3.5in]{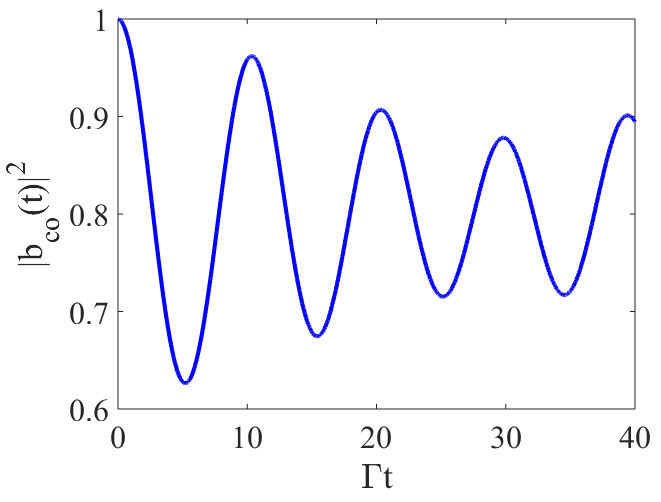}
	\end{center}
	\caption{Non-Markovian ground-state atom population dynamics in the strong-coupling regime, obtained from the numerical solution of (\ref{cg2a}). The atom is located $z_0=0.1 c/\omega_p$ above the interface, such that $g=0.808$.}
	\label{fig_pop_co_sc}
\end{figure}

For the weak-coupling case, $b_{go}\left(  t\right) \simeq e^{i\delta_{g}t}  $, and so $\left\vert b_{go}\left( t\right) \right\vert ^{2}\simeq 1$. The frequency shift (Appendix \ref{MAEAP}) is found to be $\delta_{g}=7.78\times10^{-4}\omega_{0}$, such that the real and imaginary parts of the population oscillate with a period of $\Gamma \textrm{T}\simeq151.55$, where, for reference,  $\Gamma$ is the decay rate of the excited atom, (\ref{Gamma}). Alternatively, in the strong-coupling case, there are Rabi oscillations as well as a frequency shift, $\delta_{g}=0.32\omega_{0}$, which leads to a period of $\Gamma \textrm{T}\simeq 0.4$. Figure \ref{fig_pop_co_sc} shows $\left\vert b_{go}\left(  t\right)\right\vert^2 $, where it can be seen that the population is strongly non-Markovian, and exhibits Rabi oscillations.

\begin{figure}[!htbp]
	\begin{center}
		\noindent \includegraphics[width=3.5in]{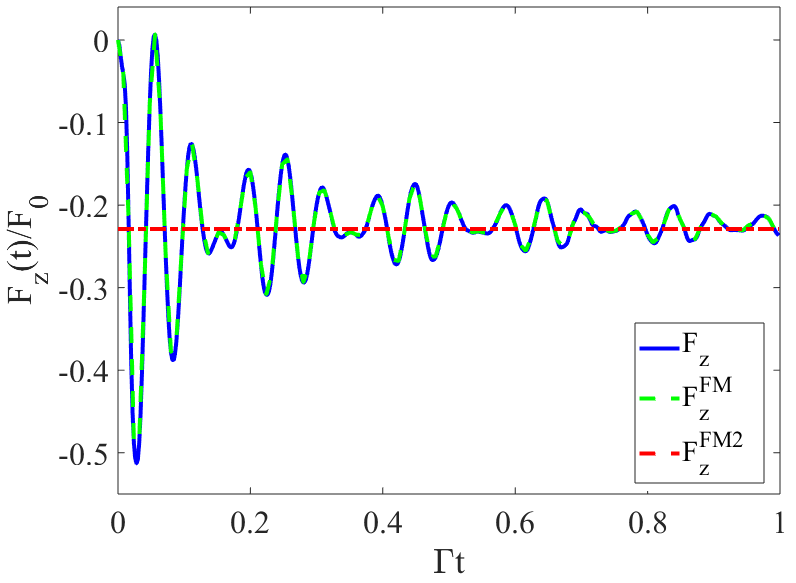}
	\end{center}
	\caption{Vertical force dynamics (transient Casimir-Polder force) on a ground-state atom in the weak-coupling regime. The atom is located $z_0=0.7 c/\omega_p$ above the interface, such that $g=0.044$.}
	\label{fig_F_PM_z}
\end{figure}

The exact, generally non-Markovian force is obtained by using the
numerically-determined amplitude $b_{go}\left(  t\right)  $ from (\ref{cg2a})
in (\ref{FG}). The vertical force (\ref{FG}) is shown in Fig. \ref{fig_F_PM_z} for the weak coupling case, along with the Markov approximation (\ref{FC42}) ((\ref{FcoPM}) is essentially the same as (\ref{FC42})) and
compared with the FM2 approximation (\ref{FC4}). We see that at $t=0$ the force has the
correct null value, and then oscillates and rapidly settles down to the value of
the FM2 approximation, which is the usual static CP force. The FM2 approximation does not have the correct value at $t=0$ due to extending upper limit of the time-integral to $t\rightarrow\infty$. Therefore, we see that the long-time ($\Gamma t\gg1$) behavior of the
vertical force on the ground-state atom is the usual Markovian Casimir-Polder
force.

The reasons for the oscillations in Fig. \ref{fig_F_PM_z} are as follows. Since we take a
bare-state, rather than dressed-state, approach, the initial state $\left\vert
g,0\right\rangle $ is not the true ground state of the atom (and, certainly,
neither is $\left\vert e,1\right\rangle $). As such, the (light-matter)
interaction can push the atom to other states, but with time the system
finally settles down into a final state that locally approximates the true ground state.

\begin{figure}[!htbp]
	\begin{center}
		\noindent \includegraphics[width=3.5in]{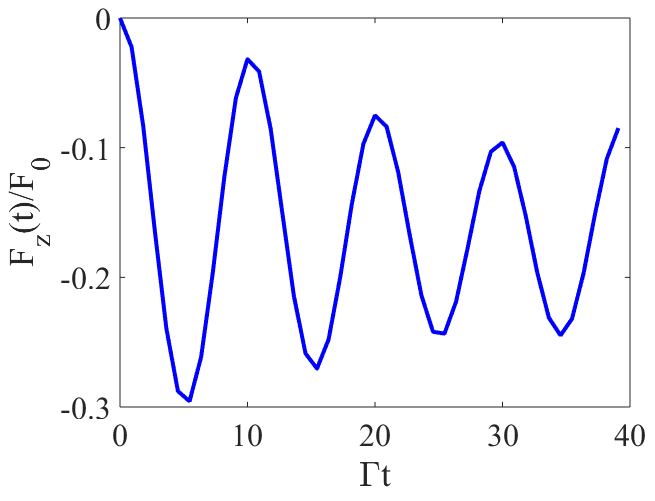}
	\end{center}
	\caption{Vertical force dynamics (transient Casimir-Polder force) on a ground-state atom in the strong-coupling regime ($z_0=0.1 c/\omega_p$, $g=0.808$).}
	\label{fig_FVert_sc}
\end{figure}

The vertical force in the strong-coupling regime is shown in Fig. \ref{fig_FVert_sc}. The strong oscillations in the force are due to the Rabi oscillations of the population. 

\subsubsection{Non-Markovian Casimir-Polder Force for $t\rightarrow\infty$ on
a Ground-State Atom\label{TII}}

In the previous section, the non-Markovian population and CP force on a ground-state atom in a
non-reciprocal structured reservoir was determined numerically (and a Markov approximation is provided in Appendix \ref{MAEAP}). Next, we consider the exact $t\rightarrow\infty$ behavior of the population and
force on a ground-state (direct-product ground state) atom. This leads to a method to
quantify the level of the non-Markovian behavior.

Starting with the energy-non-conserving states associated with $\left\vert
g,0\right\rangle $, the non-Markovian population obeys (\ref{cg2a}%
)-(\ref{cg2}), which have a convolution form. Taking Laplace transforms,%
\begin{align}
&  sb_{go}\left(  s\right)  -b_{go}\left(  t=0^{+}\right)  \\
&  =-\frac{1}{\hslash\pi\varepsilon_{0}}\int_{0}^{\infty}d\omega_{\lambda
}\ \frac{\omega_{\lambda}^{2}}{c^{2}}\frac{\mathbf{\gamma}\cdot\mathbf{G}%
_{\mathbf{I}}(\mathbf{r}_{0},\mathbf{r}_{0},\omega_{\lambda})\cdot
\mathbf{\gamma}}{2i}b_{go}\left(  s\right)  \mathcal{L}, \nonumber
\end{align}
where
\begin{align}
\mathcal{L}=\mathcal{L}\left\{  e^{-i\left(  \omega_{\lambda}+\omega_{0}\right)  t}\right\}   &
=\int_{0}^{\infty}e^{-i\left(  \omega_{\lambda}+\omega_{0}\right)  t}%
e^{-st}dt\\
&  =\frac{1}{s+i\left(  \omega_{\lambda}+\omega_{0}\right)  }.\nonumber
\end{align}
Therefore,
\begin{equation}
b_{go}\left(  s\right)  =\frac{b_{go}\left(  t=0^{+}\right)  }{s+\Gamma\left(
s\right)  }=\frac{1}{s+\Gamma\left(  s\right)  },
\end{equation}
where%
\begin{equation}
\Gamma\left(  s\right)  =\frac{1}{\hslash\pi\varepsilon_{0}}\int_{0}^{\infty
}d\omega_{\lambda}\ \frac{\omega_{\lambda}^{2}}{c^{2}}\frac{\mathbf{\gamma
}\cdot\mathbf{G}_{\mathbf{I}}(\mathbf{r}_{0},\mathbf{r}_{0},\omega_{\lambda
})\cdot\mathbf{\gamma}}{2i}\mathcal{L}.
\end{equation}

Replacing $s\rightarrow s^{\prime}-i\omega_{0}$,
\begin{equation}
b_{go}\left(  t\right)  =\frac{e^{-i\omega_{0}t}}{2\pi i}\int_{\delta-i\infty
}^{\delta+i\infty}\frac{1}{s^{\prime}-i\omega_{0}+G\left(  s^{\prime}\right)
}e^{s^{\prime}t}ds^{\prime},
\end{equation}
where $G\left(  s^{\prime}\right)=\Gamma\left(  s^{\prime}-i\omega_{0}\right)$. It can be seen that $G\left(  s^{\prime}\right)  $ has logarithmic-type branch
points at $s^{\prime}=0$ and $s^{\prime}=-i\infty$. To see that a branch cut
(BC) exists from $s^{\prime}=0$ to $s^{\prime}=-i\infty$, we can consider
\cite{BF1}-\cite{BF2} $G_{d}\left(  s^{\prime}\right)  =G\left(  x+iy\right)
-G\left(  -x+iy\right)  $,
\begin{align}
&  \lim_{x\rightarrow0}G_{d}\left(  s^{\prime}\right)  \\
&  =\frac{2}{\hslash\varepsilon_{0}}\int_{0}^{\infty}d\omega_{\lambda}%
\ \frac{\omega_{\lambda}^{2}}{c^{2}}\frac{\mathbf{\gamma}\cdot\mathbf{G}%
_{\mathbf{I}}(\mathbf{r}_{0},\mathbf{r}_{0},\omega_{\lambda})\cdot
\mathbf{\gamma}}{2i}\delta\left(  y+\omega_{\lambda}\right)  .\nonumber
\end{align}
using%
\begin{equation}
\delta\left(  y\right)  =\frac{1}{\pi}\lim_{x\rightarrow0}\frac{x}{y^{2}%
+x^{2}}.
\end{equation}
For $y>0$, the delta function is never encountered, and so $G_{d}\left(
y\right)  =0$, and there is no discontinuity. But, for $y<0$, the delta
function is encountered, and so the branch cut goes from $s^{\prime}=0$ to
$s^{\prime}=-i\infty$. Since $s=s^{\prime}-i\omega_{0}$, the BC goes from
$s=-i\omega_{0}$ to $s=-i\infty$. 

Poles will occur at $s+\Gamma\left(  s\right)=0$. For the numerical parameters assumed in Section \ref{NSR}, it is found that
there is one pole, located on the imaginary axis at $s_{p}=i\alpha_{p}$, and
$\alpha_{p}/\omega_{0}\ll1$.

\begin{figure}[!htbp]
	\begin{center}
		\noindent \includegraphics[width=3.5in]{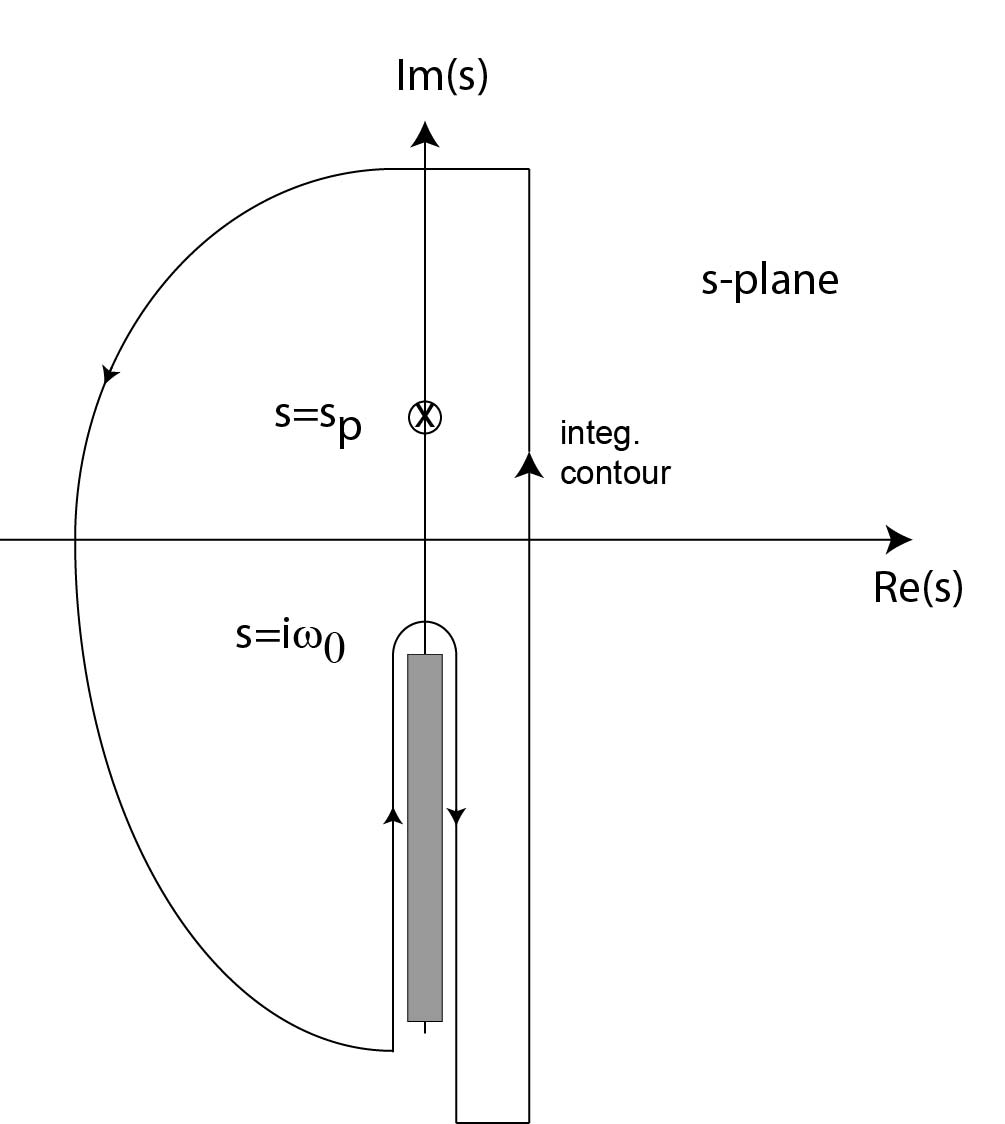}
	\end{center}
	\caption{Depiction of the $s$-plane, showing the pole, branch cut, and integration contour.}
	\label{fig_s_plane}
\end{figure}

The complex $s-$plane is depicted in Fig. \ref{fig_s_plane}, showing that the inverse Laplace
transform will involve a residue and a branch-cut integral.
\begin{align}
b_{go}\left(  t\right)   &  =\frac{1}{2\pi i}\int_{\delta-i\infty}%
^{\delta+i\infty}b_{go}\left(  s\right)  e^{st}ds\\
&  =b_{go}^{\text{Res}}\left(  t\right)  +\frac{1}{2\pi i}\int_{\text{BC}%
}\frac{1}{s+\Gamma\left(  s\right)  }e^{st}ds,
\end{align}
where%
\begin{equation}
b_{go}^{\text{Res}}\left(  t\right)  =\frac{2\pi i}{2\pi i}\frac{1}%
{\Gamma^{\prime}\left(  s_{p}\right)  }e^{s_{p}t}=ce^{i\alpha_{p}t}. \label{ap}
\end{equation}

By the Riemann-Lebesgue lemma, the branch-cut contribution goes to zero as
$t\rightarrow\infty$, so that $b_{go}\left(  t\rightarrow\infty\right)
=b_{go}^{\text{Res}}\left(  t\right)  $. This can be compared to (\ref{bgoFM1}), with the difference being the value of the oscillation frequency,
$\delta_{g}$ in (\ref{bgoFM1}) and $\alpha_{p}$ in (\ref{ap}).

Having considered the population, we want to evaluate the $t\rightarrow\infty$
value of the force (\ref{FG}). The method of directly evaluating this using
Laplace transforms is cumbersome, and so we will, instead, insert the
population obtained above, $\lim_{t\rightarrow\infty}b_{go}\left(  t\right)
=b_{go}^{\text{Res}}\left(  t\right)  =ce^{i\alpha_{p}t}$, into (\ref{FG}),
leading to%
\begin{align}
&  \mathcal{FC}_{\alpha} \label{FCNMU} \\ 
&  =\left\vert c\right\vert ^{2}2\operatorname{Re}\left\{  \frac{i}%
{\pi\varepsilon_{0}}\int_{0}^{\infty}d\omega_{\lambda}\ \frac{\omega_{\lambda
}^{2}}{c^{2}}\left.  \frac{\partial}{\partial\alpha}\frac{\mathbf{\gamma}%
\cdot\mathbf{G}_{\mathbf{I}}\left(  \mathbf{r},\mathbf{r}_{0},\omega_{\lambda
}\right)  \cdot\mathbf{\gamma}}{2i}\right\vert \right.  _{\mathbf{r}%
=\mathbf{r}_{0}}\nonumber\\
&
\ \ \ \ \ \ \ \ \ \ \ \ \ \ \ \ \ \ \ \ \ \ \ \ \ \ \ \ \ \ \ \ \ \ \ \ \ \ \ \ \ \ \ \ \ \ \ \times
\left.  \frac{1}{\omega_{\lambda}+\omega_{0}+\alpha_{p}}\right\}  .\nonumber
\end{align}
Therefore, there is only a non-resonant component of the exact non-Markovian
Casimir-Polder force on the ground-state atom.

Comparing with the FM approximation obtained by the same method, (\ref{FC42}),
in the $t\rightarrow\infty$ limit,%
\begin{align}
\mathcal{F}_{\alpha}^{\text{FM}} &  =2\operatorname{Re}\left\{  \frac{i}%
{\pi\varepsilon_{0}}\int_{0}^{\infty}d\omega_{\lambda}\ \frac{\omega_{\lambda
}^{2}}{c^{2}}\frac{\partial}{\partial\alpha}\frac{\mathbf{\gamma}%
\cdot\mathbf{G}_{\mathbf{I}}\left(  \mathbf{r},\mathbf{r}_{0},\omega_{\lambda
}\right)  \cdot\mathbf{\gamma}}{2i}\right.  \nonumber\\
&
\ \ \ \ \ \ \ \ \ \ \ \ \ \ \ \ \ \ \ \ \ \ \ \ \ \ \ \ \ \ \ \ \times
\left.  \frac{1}{\omega_{\lambda}+\omega_{0}+\delta_{g}}\right\}  , \label{NMGS}
\end{align}
we see that if $\left\vert c\right\vert =1$ and $\alpha_{p}=\delta_{g}$ (the Lamb shift), then
these are the same. The occurrence of $\alpha_{p}\neq\delta_{g}$ and
$\left\vert c\right\vert \neq1$ differentiates the Markov and non-Markov solutions.

Numerically, for weak coupling ($z_0=0.7 c/\omega_p$) at $\omega_{0}=0.65\omega_{p}$, $\alpha
_{p}=7.80\times10^{-4}\omega_{0}$, which agrees with the frequency shift
FM\ approximation, $\delta_{g}=7.78\times10^{-4}\omega_{0}$.
Furthermore,
\begin{equation}
\left\vert c\right\vert =\left\vert b_{go}^{\text{Res}}\left(  t\right)
\right\vert =\frac{1}{\left\vert D^{\prime}\left(  s_{p}\right)  \right\vert
}=0.9995,
\end{equation}
so we have, for the pole, $\alpha_{p}\simeq\delta_{g}\ll\omega_{0}$ and
$\left\vert c\right\vert \simeq1$, in which case the non-Markovian
$t\rightarrow\infty$ result (\ref{FCNMU}) is approximately the same as the FM
result for $t\rightarrow\infty$, (\ref{NMGS}), as expected for weak coupling. For the strong-coupling case ($z_0=0.1 c/\omega_p$) at $\omega_{0}=0.65\omega_{p}$, $\alpha
_{p}=0.282\omega_{0}$, whereas the frequency shift FM approximation gives $\delta_{g}=0.3198\omega_{0}$, and 
\begin{equation}
\left\vert c\right\vert =\left\vert b_{go}^{\text{Res}}\left(  t\right)
\right\vert =\frac{1}{\left\vert D^{\prime}\left(  s_{p}\right)  \right\vert
}=0.893.
\end{equation}
As expected, $\alpha_{p}\neq\delta_{g}$ and
$\left\vert c\right\vert \neq1$ for the strongly non-Markovian case.

\begin{figure}[!htbp]
	\begin{center}
		\noindent \includegraphics[width=3.5in]{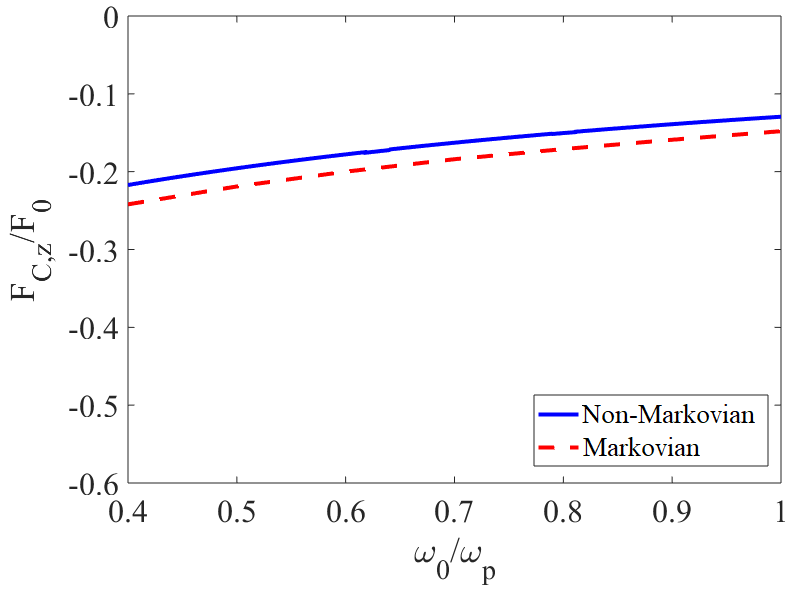}
	\end{center}
	\caption{Casimir force ($t\rightarrow\infty$) comparing the Markov and non-Markovian results for the weak-coupling situation, $z_0=0.7 c/\omega_p$, $g=0.044$.}
	\label{fig_cf}
\end{figure}

Figure \ref{fig_cf} shows the non-Markovian Casimir-Polder force ($t\rightarrow\infty$)
obtained from the residue leading to (\ref{FCNMU}), and the FM approximation, in the weak-coupling case. It
can be seen that the agreement, and the trend, agree fairly well. 

\section{Conclusions}

The non-Markovian time-dynamics of two-level atoms immersed in inhomogeneous, non-reciprocal environments has been studied using Weisskopf-Wigner theory in the strong and weak coupling regimes. Ground-state and excited atoms were considered as two
separate initial-value problems. For atoms close to a material interface, strong coupling results in strongly non-Markovian behavior. Various approximations were also discussed, and the transient Casimir-Polder force was obtained.

Our analysis reveals that the standard Markovian-type formulas used to predict the instantaneous fluctuation induced (Casimir-Polder) forces in atomic systems can be inaccurate as they neglect transients where the force can switch sign and exhibit strong oscillations. This effect is especially important in the strong coupling regime, where the usual theory totally breaks down. Furthermore, we have highlighted that the states $\left\vert
e,0\right\rangle$ and $\left\vert g,0\right\rangle$ are projected onto orthogonal subspaces of the interacting light-matter system, and thereby their time evolution is determined by two orthogonal bases of product states.


\section*{Acknowledgments}

The authors gratefully acknowledge discussions with Steve Hughes and Peter Milonni.

	\appendix
	
	\section{Green Function\label{A_GF}}
	Although the treatment is fully quantum at a macroscopic level, the needed Green function is the classical Green function, arising from classical Maxwell's equations, and is provided in \cite{PRB2018}-\cite{QT} (however, the notation
for the Green function here differs from that used in \cite{PRB2018}-\cite{QT} by a factor of $i\omega\mu_{0}$). The Green function has vacuum and scattered contributions, where the vacuum term, divergent in the dipole approximation,
leads to the Lamb shift. We assume that the Lamb shift is accounted for in the definition of the atomic transition frequency $\omega_{0}$, and in the
following we use the scattered Green function, which dominates the material response for close atom-interface separations. 
The relationship between the electric field and the Green function is \cite{Welsch0}-\cite{Phil} 
\begin{equation}
\mathbf{E}=i\omega \mu \int d^{3}\mathbf{r}^{\prime }\mathbf{G}\left( 
\mathbf{r},\mathbf{r}^{\prime },\omega \right) \cdot \mathbf{J}_{N}\left( 
\mathbf{r}^{\prime },\omega \right) ,
\end{equation}%
where 
\begin{equation}
\widehat{\mathbf{J}}_{N}\left( \mathbf{r},\omega \right) =\omega \sqrt{\frac{%
\hslash \varepsilon _{0}}{\pi }}\mathbf{T}\left( \mathbf{r},\omega \right)
\cdot \widehat{\mathbf{f}}\left( \mathbf{r},\omega \right)   \label{jnnr}
\end{equation}%
is the noise current, $\hat{\boldsymbol{\mathrm{f}}}%
,\hat{\boldsymbol{\mathrm{f}}}^{\dag}$ are the canonically conjugate field
variables, and where $\boldsymbol{\mathrm{T}}(\mathbf{r},\omega_{\lambda})\cdot
\boldsymbol{\mathrm{T}}^{\dag}(\mathbf{r},\omega_{\lambda})=\frac{1}%
{2i}\left(  \mathbf{\varepsilon}\left(  \mathbf{r},\omega\right)
-\mathbf{\varepsilon}^{\dag}\left(  \mathbf{r},\omega\right)  \right)  $ accounts for the material environment.

The Green function satisfies \cite{Hanson}
\begin{align}
\text{ }2i\frac{\omega^{2}}{c^{2}}&\int d^{3}\mathrm{r}^{\prime}%
K_{ik}\left(  \mathbf{r},\mathbf{r}^{\prime},\omega_{\lambda}\right)
K_{jk}^{\ast}\left(  \mathbf{r}_{0},\mathbf{r}^{\prime},\omega_{\lambda
}\right)  \label{magic}\\
&  =G_{ij}(\mathbf{r},\mathbf{r}_{0},\omega_{\lambda})-G_{ij}^{\ast
}(\mathbf{r}_{0},\mathbf{r},\omega_{\lambda})=G_{\mathbf{I},i,j}\left(
\mathbf{r},\mathbf{r}_{0},\omega_{\lambda}\right)  .\nonumber
\end{align}
Writing a scalar component of the Green function as
\begin{align}
G(\mathbf{r},\mathbf{r}_{0},\omega) &  \sim\int dk_{x}dk_{y}\ \left(
\widetilde{G}_{r}\left(  k_{x},k_{y}\right)  +i\widetilde{G}_{i}\left(
k_{x},k_{y}\right)  \right) \nonumber \\
&\times e^{ik_{x}\left(  x-x_{0}\right)  }e^{ik_{y}\left(
y-y_{0}\right)  }e^{-\gamma_{0}\left(  k_{x},k_{y}\right)  \left(
z+z_{0}\right)  },
\end{align}
for the layered environment depicted in Fig. \ref{Fig_geom}, where $\gamma_0 = \sqrt{{k_x}^2+{k_y}^2-{k_0}^2 }$, it is easily shown that
\begin{align}
\frac{\partial}{\partial\alpha} &  \frac{\mathbf{\gamma}\cdot\mathbf{G}%
_{\mathbf{I}}\left(  \mathbf{r},\mathbf{r}_{0},\omega\right)  \cdot
\mathbf{\gamma}}{2i}\label{DXZ}\\
&  =\left\{
\begin{array}
[c]{c}%
-i\operatorname{Re}\frac{\partial}{\partial\alpha}\mathbf{\gamma}%
\cdot\mathbf{G}\left(  \mathbf{r}_{0},\mathbf{r}_{0},\omega\right)
\cdot\mathbf{\gamma},\ \ \ \alpha=x,y\\
\operatorname{Im}\frac{\partial}{\partial\alpha}\mathbf{\gamma}\cdot
\mathbf{G}\left(  \mathbf{r}_{0},\mathbf{r}_{0},\omega\right)  \cdot
\mathbf{\gamma},\ \ \ \alpha=z.
\end{array}
\right.  \nonumber
\end{align}

	\section{Numerical Solution of Volterra Integral Equation\label{NS}}
	
	In order to numerically solve the Volterra integral equation (\ref{VE1}) having the form 
\begin{equation}
\frac{d}{dt}c\left(  t\right)  =\int_{0}^{t}H\left(  t,t^{\prime}\right)
c\left(  t^{\prime}\right)  dt^{\prime},
\end{equation}
a grid can be defined \cite{NR} $t_{i}=0+ih,\ \ \ i=0,1,...N$, where $h=t_{\text{final}}/N$ with $N$ the number of grid points
($t_{\text{initial}}=0$ is implicit), and using a trapezoidal rule
\begin{align}
&  \int_{0}^{t_{i}}H\left(  t_{i},t^{\prime}\right)  c\left(  t^{\prime
}\right)  dt^{\prime}\\
&  =h\left(  \frac{1}{2}H_{i0}c_{0}+\sum_{j=1}^{i-1}H_{ij}c_{j}+\frac{1}%
{2}H_{ii}c_{i}\right)  ,\nonumber
\end{align}
where $H_{ij}=H\left(  t_{i},t_{j}\right)  $, $c_{j}=c\left(
t_{j}\right)  $. Writing the derivative as
\begin{equation}
\frac{d}{dt}c\left(  t\right)  =\frac{c\left(  t+h\right)
-c\left(  t\right)  }{h},
\end{equation}
then,
\[
\frac{c_{\left(  i+1\right)  }-c_{i}}{h}-h\frac{1}{2}H_{ii}c_{i}%
=h\left(  \frac{1}{2}H_{i0}c_{0}+\sum_{j=1}^{i-1}H_{ij}c_{j}\right)  ,
\]
$i=0,1,2,...N$, where for $i=0$, $c_{0}=1$. In general,%
\begin{align}
c_{m}= &  \left(  1+h^{2}\frac{1}{2}H_{\left(  m-1\right)  \left(
m-1\right)  }\right)  c_{\left(  m-1\right)  }\\
+ &  h^{2}\left(  \frac{1}{2}H_{\left(  m-1\right)  0}+\sum_{j=1}%
^{m-2}H_{\left(  m-1\right)  j}c_{j}\right)  ,\ \ \ m=1,2,...N.\nonumber
\end{align}

\section{Markov Approximations of the Population\label{MAEAP}}

\subsection{Excited Atom}
Various Markov-type approximations can be made in evaluating the time integral in
(\ref{VE1}) for the weak coupling case, where the result is essentially Markovian. The first approximation is to assume that the population has no memory (Markov
approximation, MA), $c_{eo}\left(  t^{\prime}\right)  \simeq c_{eo}\left(
t\right)  $, and the second approximation is to extend the upper limit of the
integration to infinity, which can typically be justified by noting that the
most important contribution to the integral comes from the vicinity of
$\omega_{\lambda}=\omega_{0}$. Then, the Sokhotski--Plemelj (SP) identity,
\begin{align}
&  \int_{0}^{t}e^{\pm i\left(  \omega-\omega_{0}\right)  \left(
t-t^{\prime}\right)  }dt^{\prime} \label{SPI} \\
&  \rightarrow\int_{0}^{\infty}e^{\pm i\left(  \omega-\omega
_{0}\right)  \left(  t-t^{\prime}\right)  }dt^{\prime}=\pi\delta\left(
\omega-\omega_{0}\right)  \pm i\text{PV}\left(  \frac{1}{\omega-\omega_{0}%
}\right)  \nonumber
\end{align}
leads to the usual resonant and non-resonant contributions. Since these two
approximations are often used together, we will refer to this as the full
Markov (FM) approximation.

Another option is to assume that the population has no memory (MA), but that the upper
limit of the integration is not extended to infinity, leading to
\begin{equation}
\int_{0}^{t}e^{-i\left(  \omega-\omega_{0}\right)  \left(  t-t^{\prime
}\right)  }dt^{\prime}=\frac{1-e^{-i\left(  \omega-\omega_{0}\right)  t}%
}{i\left(  \omega-\omega_{0}\right)  }.\label{PM}%
\end{equation}
We will refer to this as the partial Markov (PM) approximation. In the
following it will be useful to refer to the function
\begin{equation}
h_{e}\left(  \mathbf{r},\mathbf{r},\omega,g\right)  =\frac{1}{\hslash
\pi\varepsilon_{0}}\int_{0}^{\infty}d\omega_{\lambda}\ \frac{\omega_{\lambda
}^{2}}{c^{2}}\frac{\mathbf{\gamma}\cdot\mathbf{G}_{\mathbf{I}}(\mathbf{r}%
,\mathbf{r},\omega_{\lambda})\cdot\mathbf{\gamma}}{2i(\omega_{\lambda}-\omega)}g,\label{hef}%
\end{equation}
where $g=g(\omega_{\lambda},t)$.

The FM approximation of the Volterra integral equation yields
\begin{equation}
\frac{d}{dt}c_{eo}^{\text{FM}}\left(  t\right) =\left(  -\Gamma^{\text{FM}}\frac{1}{2}+i\delta^{\text{FM}}\right)  c_{eo}^{\text{FM}}\left(  t\right)
,\label{dce2}%
\end{equation}
with energy shift is $\delta^{\text{FM}}=\delta=\text{PV} \left( h_{e}\left(  \mathbf{r_0},\mathbf{r_0},\omega_0 ,1\right) \right)$, where PV indicates a principal-value integral, and decay rate $\Gamma^{\text{FM}}$ 
\begin{equation}
\Gamma^{\text{FM}}  =\Gamma=\frac{2\omega_{0}^{2}}{\hslash\varepsilon
_{0}c^{2}}\frac{\mathbf{\gamma}\cdot\mathbf{G}_{\mathbf{I}}(\mathbf{r}%
_{0},\mathbf{r}_{0},\omega_{0})\cdot\mathbf{\gamma}}{2i}.\label{Gamma}
\end{equation}
Since, $\mathbf{\gamma}\cdot\mathbf{G}_{\mathbf{I}}(\mathbf{r}%
_{0},\mathbf{r}_{0},\omega_{0})\cdot\mathbf{\gamma}=2i\operatorname{Im}\mathbf{\gamma}\cdot\mathbf{G}(\mathbf{r}%
_{0},\mathbf{r}_{0},\omega_{0})\cdot\mathbf{\gamma}$ for a linear dipole, $\Gamma$ and $\delta$ are seen to be real-valued, as required,
and provide the usual exponential decay and energy shift of
$-\hslash\delta$, which agree with the well-known expressions for reciprocal
media \cite{BookVW}. Therefore, for a linear dipole the form of $\Gamma$ and $\delta$ in terms of
the Green function are the same in the reciprocal and non-reciprocal case.

From (\ref{dce2}), $c_{eo}^{\text{FM}}\left(  t\right)  =c_{eo}^{\text{FM}}\left(
0\right)  e^{-\Gamma\frac{1}{2}t}e^{i\delta t}$, such that the FM amplitude of
the state $\left\vert e,0\right\rangle $ is
\begin{equation}
b_{eo}^{\text{FM}}\left(  t\right)  =c_{eo}^{\text{FM}}\left(  0\right)  e^{-\Gamma
\frac{1}{2}t}e^{-i\left(  \omega_{0}-\delta\right)  t},\label{beoFM}%
\end{equation}
with $c_{eo}^{\text{FM}}\left(  0\right)  =1$ by assumption of the initial-value condition.

In the PM approximation,%
\begin{equation}
\frac{d}{dt}c_{eo}^{\text{PM}}\left(  t\right)  \simeq-c_{eo}^{\text{PM}%
}\left(  t\right)  p_{e}\left(  t\right),  \label{PMa1}%
\end{equation}
where $p_{e}\left(  t\right)  = \text{PV} \left( h_{e}\left(  \mathbf{r_0},\mathbf{r_0},\omega_0, f_{e}\left(  t\right) \right) \right), \ f_{e}\left(  t\right)  =-i\left(  1-e^{-i\left(  \omega_{\lambda}-\omega
_{0}\right)  t}\right)  $.
The solution of (\ref{PMa1}) is%
\begin{equation}
c_{eo}^{\text{PM}}\left(  t\right)  =e^{ih_{e}t}e^{q_{e}\left(  t\right)
-q_{e}\left(  0\right)  } \label{sPM1},
\end{equation}
where $q_{e}\left(  t\right)  =\text{PV} \left( h_{e}\left(  \mathbf{r_0},\mathbf{r_0},\omega_0, g_{e}\left(  t\right) \right) \right), \  g_{e}\left(  t\right)  =e^{-i\left(  \omega_{\lambda}-\omega_{0}\right)
t}/\left(  \omega_{\lambda}-\omega_{0}\right)  $.

Since by causality $\mathbf{G}_{\mathbf{I}}(\mathbf{r}_{0},\mathbf{r}%
_{0},\omega_{\lambda})$ must be analytic in the upper-half $\omega_{\lambda}%
$-plane, and $\lim_{\left\vert \omega_{\lambda}\right\vert \rightarrow\infty
}\left(  \omega_{\lambda}^{2}/c^{2}\right)  \mathbf{G}_{\mathbf{I}}%
(\mathbf{r},\mathbf{r}_{0},\omega_{\lambda})=0$, the integral for $h_{e}$ can
be closed with a semi-circle in the first quadrant of the complex
$\omega_{\lambda}$-plane, resulting in an integral over positive imaginary
frequencies. 

\subsection{Ground-State Atom}

For the ground-state atom, in the PM approximation of (\ref{cg2a}),%
\begin{equation}
\frac{d}{dt}b_{go}^{\text{PM}}\left(  t\right)  =-b_{go}^{\text{PM}}\left(
t\right)  p_{g}\left(  t\right)  .\label{bgopm}%
\end{equation}
where $p_{g}\left(  t\right)  = h_{e}\left(  \mathbf{r_0},\mathbf{r_0},-\omega_0, f_{g}\left(  t\right) \right), \, f_{g}\left(  t\right)  =-i\left(  1-e^{-i\left(  \omega_{\lambda}+\omega
_{0}\right)  t}\right)  $. The solution of (\ref{bgopm}) is%
\begin{equation}
b_{go}^{\text{PM}}\left(  t\right)  =e^{i\delta_{g}t}e^{q_{g}\left(  t\right)
-q_{g}\left(  0\right)  },
\end{equation}
where  $\delta_{g} = h_{e}\left(  \mathbf{r_0},\mathbf{r_0},-\omega_0, 1 \right), \ q_{g}\left(  t\right)  = h_{e}\left(  \mathbf{r_0},\mathbf{r_0},-\omega_0, r_{g}\left(  t\right) \right)$, and $r_{g}\left(  t\right)  =e^{-i\left(  \omega_{\lambda}+\omega_{0}\right) t}/\left(  \omega_{\lambda}+\omega_{0}\right)  $.

It can be seen that $q_{g}\left(  t\right)  $ rapidly becomes small as $t$
increases, due to the rapidly-oscillating integrand, and so
\begin{equation}
b_{go}^{\text{PM}}\left(  t\right)  \simeq e^{i\delta_{g}t},\label{bPM}%
\end{equation}
which agrees with the result from the full Markov approximation, $\frac{d}%
{dt}b_{go}^\text{FM}\left(  t\right)  =b_{go}^\text{FM}\left(  t\right)  i\delta_{g},$ so that
\begin{equation}
b_{go}^{\text{FM}}\left(  t\right)  =e^{i\delta_{g}t}.\label{bgoFM1}%
\end{equation}
Therefore, in the Markov approximation, the state $\left\vert g,0\right\rangle
$ has no decay \cite{Mil}, unlike the state $\left\vert e,0\right\rangle $.
The relative energy difference between the states $\left\vert e,1\right\rangle
$ and $\left\vert g,0\right\rangle $ is $\hslash\omega_{0}$.
\section{Markov Approximations of the Force\label{MAEA}}

\subsection{Excited Atom}

The exact, generally non-Markovian force is given by (\ref{F2}). There are several combinations of Markov-type approximations that can be used to approximate the force in the weak coupling case. One form of
Markov approximation of the force is obtained by substituting the PM or FM
approximations for the population into the force expression, then evaluating
the resulting time-integral exactly. In this manner, for example, the resulting FM approximation of
the force is
\begin{align}
&  \mathcal{F}_{\alpha}^{\text{FM}}\left(  t\right)  \label{FFM}\\
&  =2\operatorname{Re}\left\{  \frac{i}{\pi\varepsilon_{0}}\int_{0}^{\infty
}d\omega_{\lambda}\frac{\omega_{\lambda}^{2}}{c^{2}}\left.  \frac{\partial
}{\partial\alpha}\frac{\mathbf{\gamma}\cdot\mathbf{G}_{\mathbf{I}}\left(
\mathbf{r},\mathbf{r}_{0},\omega_{\lambda}\right)  \cdot\mathbf{\gamma}}%
{2i}\right\vert _{\mathbf{r}=\mathbf{r}_{0}}\right.  \nonumber\\
&
\ \ \ \ \ \ \ \ \ \ \ \ \ \ \ \ \ \ \ \ \ \ \ \ \ \times
\left.  e^{-\Gamma\frac{1}{2}t}\frac{e^{-i\left(  \omega_{\lambda}-\omega
_{0}+\delta\right)  t}-e^{-\Gamma\frac{1}{2}t}}{\Gamma-i\left(  \omega
_{\lambda}-\omega_{0}+\delta\right)  }\right\},  \nonumber
\end{align}
and similarly for $\mathcal{F}_{\alpha}^{\text{PM}}$. Alternatively, one could first impose the Markov approximation $c_{eo}\left(
t^{\prime}\right)  \simeq c_{eo}\left(  t\right)  $ directly in the time integral in
(\ref{F2}), then evaluating the resulting time-integral exactly, resulting in 

\begin{align}
&  \mathcal{F}_{\alpha}^{\text{PM2}}\left(  t\right)  \simeq\frac{2}%
{\pi\varepsilon_{0}}\left\vert c_{eo}\left(  t\right)  \right\vert
^{2}\label{MF}\\
&  \times\operatorname{Re}\left\{  \int_{0}^{\infty}d\omega_{\lambda}%
\frac{\omega_{\lambda}^{2}}{c^{2}}\left.  \frac{\partial}{\partial\alpha}%
\frac{\mathbf{\gamma}\cdot\mathbf{G}_{\mathbf{I}}\left(  \mathbf{r}%
,\mathbf{r}_{0},\omega_{\lambda}\right)  \cdot\mathbf{\gamma}}{2i}\right\vert
_{\mathbf{r}=\mathbf{r}_{0}}\right.  \nonumber\\
&
\ \ \ \ \ \ \ \ \ \ \ \ \ \ \ \ \ \ \ \ \ \ \ \ \ \ \ \ \ \ \ \ \ \ \ \ \ \ \ \ \times
\left.  \frac{1-e^{-i\left(  \omega_{\lambda}-\omega_{0}\right)  t}}{\left(
\omega_{\lambda}-\omega_{0}\right)  }\right\}  .\nonumber
\end{align}
As a further approximation, the upper-limit of the time-integral could be extended to $t\rightarrow\infty$,
allowing the SP identity to be used. However, this leads to non-zero force at $t=0$.

In a Markovian approximation, the Casimir-Polder force can be obtained as a derivative of the
Markovian energy shift, $\hslash \delta^{\text{FM}}=\hslash \text{PV} \left(h_{e}\left(  \mathbf{r},\mathbf{r},\omega_0 ,1\right) \right)$, which is the same as \cite[(4.39), (6.77)]{B2}. There, they assume an excited atom, akin to starting with the
state $\left\vert e,0\right\rangle $. The CP force can then
be written as the total differential of the energy shift,
\begin{align}
\mathcal{FC}_{z} &  =-d\left(  -\hslash\delta\left(  \mathbf{r}\right)
\right)  \label{FES1}\\
&  =\frac{2}{\pi\varepsilon_{0}}\text{PV}\int_{0}^{\infty}d\omega_{\lambda
}\ \frac{\omega_{\lambda}^{2}}{c^{2}}\left.  \frac{\partial}{\partial z}%
\frac{\mathbf{\gamma}\cdot\mathbf{G}_{\mathbf{I}}(\mathbf{r},\mathbf{r}%
_{0},\omega_{\lambda})\cdot\mathbf{\gamma}}{2i\left(  \omega_{\lambda}%
-\omega_{0}\right)  }\right\vert _{\mathbf{r}=\mathbf{r}_{0}}.\nonumber
\end{align}
Using the Wick rotation, complex-plane
analysis leads to resonant and nonresonant components.

\subsection{Ground-State Atom}
For the force on a ground-state atom, (\ref{FG}), if one first evaluates the PM or FM
population, $b_{go}^{\text{FM}}\left(  t\right)  =e^{i\delta_{g}t}\simeq
b_{go}^{\text{PM}}\left(  t\right)  $, and inserts this into the force
equation and evaluates the time integral exactly, the result is, e.g.,  
\begin{align}
\mathcal{F}_{\alpha}^{\text{FM}}\left(  t\right)   &  =2\operatorname{Re}%
\left\{  \frac{i}{\pi\varepsilon_{0}}\int_{0}^{\infty}d\omega_{\lambda}%
\ \frac{\omega_{\lambda}^{2}}{c^{2}}\frac{\partial}{\partial\alpha}%
\frac{\mathbf{\gamma}\cdot\mathbf{G}_{\mathbf{I}}\left(  \mathbf{r}%
,\mathbf{r}_{0},\omega_{\lambda}\right)  \cdot\mathbf{\gamma}}{2i}\right.
\notag\\
&  \ \ \ \ \ \ \ \ \ \ \ \ \ \ \ \ \ \ \ \ \ \times\left.
\frac{1-e^{-i\left(  \omega_{\lambda}+\omega_{0}+\delta_{g}\right)  t}}%
{\omega_{\lambda}+\omega_{0}+\delta_{g}}\right\}  .\label{FC42}
\end{align}

An other option is to first impose the Markov approximation $b_{go}\left(
t^{\prime}\right)  \simeq b_{go}\left(  t\right)  $ in the time integral in
(\ref{FG}), and then evaluate the time-integral exactly, without extending the
upper limit to $t\rightarrow\infty$ (PM2), or extending the upper limit to
$t\rightarrow\infty$ and using the SP identity (FM2), leading to
\begin{align}
\mathcal{F}_{\alpha }^{\text{PM2}}\left( t\right) & \simeq \left\vert
b_{go}\left( t\right) \right\vert ^{2}\frac{2}{\pi \varepsilon _{0}}
\label{FcoPM} \\
& \times \left\{ \textrm{Re}\int_{0}^{\infty }d\omega _{\lambda }\frac{\omega
_{\lambda }^{2}}{c^{2}}\left. \frac{\partial }{\partial \alpha }\frac{%
\mathbf{\gamma }\cdot \mathbf{G}_{\mathbf{I}}\left( \mathbf{r},\mathbf{r}%
_{0},\omega _{\lambda }\right) \cdot \mathbf{\gamma }}{2i}\right\vert _{%
\mathbf{r}=\mathbf{r}_{0}}\right.   \nonumber \\
& \ \ \ \ \ \ \ \ \ \ \ \ \ \ \ \ \ \ \ \ \ \ \ \ \ \ \ \ \ \ \times \left. 
\frac{1-e^{-i\left( \omega _{\lambda }+\omega _{0}\right) t}}{\omega
_{\lambda }+\omega _{0}}\right\} ,  \nonumber \\
& =\left\vert b_{go}\left( t\right) \right\vert ^{2}\left( \mathcal{F}%
_{\alpha \text{, static}}^{\text{PM2}}+\mathcal{F}_{\alpha \text{, dynamic}%
}^{\text{PM2}}\left( t\right) \right)   \nonumber \\
& \simeq \mathcal{F}_{\alpha \text{, static}}^{\text{PM2}}+\mathcal{F}%
_{\alpha \text{, dynamic}}^{\text{PM2}}\left( t\right)   \nonumber \\
\mathcal{F}_{\alpha }^{\text{FM2}}\left( t\right) & \simeq \left\vert
b_{go}\left( t\right) \right\vert ^{2}\frac{2}{\pi \varepsilon _{0}}
\label{FC4} \\
\times \textrm{Re}& \left\{ \int_{0}^{\infty }d\omega _{\lambda }\ \frac{%
\omega _{\lambda }^{2}}{c^{2}}\left. \frac{\partial }{\partial \alpha }\frac{%
\mathbf{\gamma }\cdot \mathbf{G}_{\mathbf{I}}\left( \mathbf{r},\mathbf{r}%
_{0},\omega _{\lambda }\right) \cdot \mathbf{\gamma }}{2i\left( \omega
_{\lambda }+\omega _{0}\right) }\right\vert _{\mathbf{r}=\mathbf{r}%
_{0}}\right\}   \nonumber \\
& =\left\vert b_{go}\left( t\right) \right\vert ^{2}\mathcal{F}_{\alpha 
\text{, static}}^{\text{PM2}}\simeq \mathcal{F}_{\alpha \text{, static}}^{%
\text{PM2}}  \nonumber
\end{align}%
where $\left\vert b_{go}\left( t \right) \right\vert ^{2}\simeq 1$ for weak
coupling. Since $\delta _{g}\ll \omega _{0}$, $\mathcal{F}_{\alpha }^{\text{PM}}\left(
t\right) \simeq \mathcal{F}_{\alpha }^{\text{PM2}}\left( t\right) $. The static and dynamic terms $\mathcal{F}%
_{\alpha \text{, static}}^{\text{PM2}} $ and $\mathcal{F}_{\alpha \text{, dynamic}%
}^{\text{PM2}}\left( t\right) $ correspond to static and dynamic potentials that agree with \cite[(15)]{DCE3}.

The FM approximation (\ref{FC42}) and PM2 approximation (\ref{FcoPM}) agree very
well with the exact force (\ref{FG}) for the weak coupling case, since the
system is essentially Markovian. The FM2\ approximation results in $\mathcal{%
F}_{\alpha }^{\text{FM2}}\left( 0\right) \neq 0$, but for longer times, $%
\mathcal{F}_{\alpha }^{\text{FM}}\left( t\right) \simeq \mathcal{F}_{\alpha
}^{\text{PM}}\left( t\right) \simeq \mathcal{F}_{\alpha }^{\text{FM2}}\left(
t\right) \simeq \mathcal{F}_{\alpha }\left( t\right) $. Since this is the
force on the ground-state atom, this can be considered as the CP force, $%
\mathcal{F}C_{\alpha }$.

For the Casimir-Polder force, from the FM approximation of the population, $%
\hslash \delta_g=\hslash h_{e}\left( \mathbf{r},\mathbf{r},-\omega_0
,1\right) $, which is the same as \cite[(4.50)]{B1} (there they assume a
ground-state atom, which is essentially the same as starting with the state $%
\left\vert g,0\right\rangle $). Then, we can write the Casimir-Polder force
as the total differential of the energy shift, $\mathcal{FC}=-d\left(
-\hslash \delta_g \left( \mathbf{r}\right) \right) $, which is the same as (%
\ref{FC4}).


\begin{thebibliography}{99}
\bibitem {FN}The force $\mathcal{F}_{\alpha}\left(  0\right)  =0$ is expected
based on the following argument: In the Heisenberg picture all the operators
at $t=0$ are taken as the free-field operators, and the initial state is by
assumption the product of the ground states of the non-interacting
Hamiltonians. Therefore, $\left\langle \hat{\boldsymbol{\mathrm{p}}}\left(
0\right)  \cdot\partial_{\alpha}\hat{\mathbf{E}}\left(  \mathbf{r},0\right)
\right\rangle =\left\langle \hat{\boldsymbol{\mathrm{p}}}\left(  0\right)
\right\rangle \cdot\left\langle \partial_{\alpha}\hat{\mathbf{E}}\left(
\mathbf{r},0\right)  \right\rangle =0\cdot0=0$ for all components of the force.
\end{thebibliography}
\end{document}